\documentclass{aa}  

\usepackage{graphicx}
\usepackage{lipsum} 
\usepackage{float}   %
\usepackage{caption}
\usepackage{placeins}
\usepackage[switch]{lineno}
\usepackage{dblfloatfix}
\usepackage[normalem]{ulem}
\usepackage[dvipsnames]{xcolor}
\usepackage{subcaption} 
\usepackage{txfonts}
\usepackage{natbib}
\bibpunct{(}{)}{;}{a}{}{,} 
\usepackage{amsmath}
\begin{document} 

   \title{A multi-wavelength approach of AGN feedback in LINERs: The case of NGC 4438}

   \subtitle{}

   \author{ Puig-Subirà M.\inst{1}
          \and
        Moldón, J. \inst{1}
        \and
        Márquez, I. \inst{1}
        \and
        Masegosa, J. \inst{1}
        \and
        González-Martín, O. \inst{1,} \inst{2}
        \and
        Hermosa Muñoz, L. \inst{3}
        \and
        Cazzoli, S. \inst{1}
        \and
        Williams-Baldwin, D.  \inst{4}
          }

   \institute{Instituto de Astrofísica de Andalucía - CSIC, Glorieta de la Astronomía s/n, 18008 Granada, Spain\\
              \email{mpuig@iaa.es}
            \and IRyA - Instituto de Radioastronomía y Astrofísica, 3-72 Xangari, 8701, Morelia, Mexico\
            \and Centro de Astrobiología (CAB) CSIC-INTA, Camino Bajo del Castillo s/n, 28692 Villanueva de la Cañada, Madrid, Spain\
             \and Jodrell Bank Centre for Astrophysics, School of Physics and Astronomy, The University of Manchester, Alan Turing Building, Oxford Road, Manchester M13 9PL, UK}

   \date{Accepted for publication in A$\&$A, January 2026}

 
  \abstract
   {The outflows generated by active galactic nuclei (AGN) may play a crucial role in galaxy evolution. In order to better understand how the feedback from AGN works, multi-wavelength studies, including low-luminosity AGN (LLAGN), are required. In particular, the presence of multi-phase outflows in low ionisation nuclear emission-line regions (LINERs) has been confirmed to be frequent but the mechanisms that launch them are still under study.}
   {We aim to explore the connections between the ionised gas outflow, radio continuum structures and X-ray emission detected in the LINER NGC\,4438. We sought morphological and energetic evidence of jet-mode feedback, revealing how a jet may modify the ionised gas structure and determine whether the ionised gas outflow can be driven by the jet.}
  {We analyse L, C and X-band images (from 1.4 to 12 GHz) of the LINER NGC\,4438, combining high-resolution data from enhanced Multi Element Radio Linked Interferometer Network (e-MERLIN) and Karl G Jansky Very Large Array (VLA). We produce radio flux and spectral index maps from which we characterise the source. In particular, we create an energetic model that allows us to estimate the power, age, magnetic field or the velocity of the particles of the jet. We incorporate optical integral field spectroscopy (IFS) data (GTC/MEGARA) and \textit{Chandra} X-ray data, with comparable resolution, to better trace the outflow, the AGN and their potential connection.}
  {We present new L, C, and X-band high-resolution, high-sensitivity radio images and spectral-index maps that probe $\sim$25~pc scales in NGC~4438. These data reveal a close morphological correspondence between the radio structures and the ionised gas bubble. Using a spatially resolved energetic model based on radio flux and spectral index, we disentangle the compact AGN emission from the extended bubble for the first time, establishing their distinct physical origins. We measure a kinetic power of $\sim 5\times 10^{44}$\,erg\,s$^{-1}$ for the radio bubble, exceeding the power of the ionised outflow by more than three orders of magnitude.} 
  {Our multi-wavelength analysis indicates that NGC 4438 is undergoing jet-mode feedback, where a low-luminosity, weakly collimated jet impacts the dense northern interstellar medium. This interaction drives shock-ionised gas, produces a moderate-velocity outflow that removes material from the region, and generates thermal X-ray emission coincident with the radio and H$\alpha$ cavity.}


   \maketitle

\section{Introduction}\label{Intro}

Low ionisation nuclear emission-line regions (LINER) represent the most numerous active galactic nuclei (AGN) population in the local Universe \citep{H80,Ho08,OGM06,GM09,M17}. Their lack of strong star formation (SF) makes them a promising opportunity to study AGN feedback, in particular in their relative innermost regions. Winds involving gas at different phases and temperatures are detected arising from these regions. We refer to them as multi-phase outflows, considering their multi-wavelength nature - ionised, neutral, atomic or molecular gas - and the interplay between all the phases \citep{C14, F21}. \cite{HM22} confirm the frequency of multi-phase outflows in LINERs by combining morphological and kinematic criteria (see below). They find hints of outflows in at least one out of every three nearby LINERs in their H$\alpha$ atlas, which is composed of 70 sources. However, the mechanism that launches and drives outflows in radiatively inefficient systems with low Eddington ratios, such as LINERs, is still under investigation. The "kinetic or jet-mode" postulated by \cite{F12} appears as a plausible alternative to explain the feedback in these systems, where accretion discs in the form of advection-dominated accretion flows (ADAF) may be responsible of compact nuclear radio emission and jets (\citealt{N98}; \citealt{FO23}). Jets are defined as highly collimated outflows of plasma launched from compact objects, namely AGN (\citealt{B19}). This material, ejected from an accreting supermassive black hole (SMBH), emits synchrotron radiation and photons that are up-scattered to higher energies via Compton scattering (\citealt{P19}). The fraction of jets found in local, low-luminosity radio-quiet AGN, particularly in LINERs, is higher than in more powerful quasars (\citealt{P19}; \citealt{LEMMINGSI}). Thus, the commonly observed radio jets in LINERs appear as potential power sources for their outflows. Indeed, the possible relation between the jets and the outflows is suggested in other LLAGN as the LINER NGC\,1052 \citep{C22} or the Seyferts NGC\,1125 \citep{S25} and NGC\,5972 \citep{A25}. To investigate whether the radio jet can be responsible for driving the observed outflow, we propose a morphological and energetic study focused in the connection of its ionised phase and the radio continuum emission. For that purpose, an in-depth radio analysis with high-sensitivity and high-resolution data is essential.

In this paper, we focus on the LINER NGC\,4438, which is a highly inclined spiral galaxy located near the centre of the Virgo cluster (d = 17 Mpc; \citealt{Vo05}). Among Virgo galaxies, NGC\,4438 stands out as the most perturbed, showing a complex morphology, associated to high-velocity interstellar medium (ISM) collision with NGC 4435 and interaction with the dense intracluster medium in the Virgo core \citep[see e.g.][] {K83,C88,K95,M04}. Several works in the literature agree on low SF in the inner parts of the galaxy \citep[see e.g.][]{V12,M19}, that is composed by an old population with no significant contribution from any recent starburst (region 4 in \citealt{B05}). In \cite{M19}, they report a total star formation rate (SFR) for NGC\,4438 of $\sim0.50\,$M$_{\odot}$yr$^{-1}$, accounting for the whole galaxy. In the nuclear region \cite{H07} and \cite{K02} give modest SFR of $\sim 0.05 - 0.1\,$M$_{\odot}$yr$^{-1}$. Neither the detected radio emission nor the ionised gas outflow can be explained by the energy coming only from SF. Optical spectroscopic studies classify NGC\,4438 as a LINER \citep{S82,K83,H83,C18,HM22}. The AGN nature is suggested by a broad H$\alpha$\,$\lambda$\,6563\,\r{A} emission in the nucleus \citep{H97,C18} and confirmed via X-rays as a compact source at hard X-rays \citep{GM09,Li22}. The location of its core was somewhat ambiguous, partly due to a dust band that obscures the central region \citep{K02}. The asymmetric morphology of this galaxy, extending westward on both large scales and nuclear regions, makes it an appealing target for multi-wavelength and multi-scale studies. At radio frequencies, \cite{Hu83} point to an elongated extended feature with two distinct components orientated along the minor axis of the galaxy. \cite{HS91} also report this structure highlighting the asymmetry between the two blobs. 
They cannot distinguish the nucleus and instead suggest that its location resides within the main radio structure. Among the extended double-lobe structures, \cite{H07} show a compact radio source with a flatter spectrum that was proposed to be the core. From their Karl G Jansky Very Large Array (VLA) observations, they also show an L-C band spectral index map and energetic model. Both with a resolutions of 1.5 arcsecond, the substructures are blended so they can not be analysed separately.

\cite{M11} propose the H$\alpha$ bubble emerging from the nucleus, identified in Hubble Space Telescope (HST) images, as an ionised gas outflow based on their morphological analysis.
\cite{HM24} confirm it by kinematic analysis using integral field spectroscopy (IFS) data from MEGARA/GTC. BPT diagrams confirm that the ionisation of the outflow gas does not originate from SF (\citealp{L25}); rather, it would result from AGN or shocks, which is consistent with its LINER nature (see \citealp{M17}; \citealp{C18}). The ionised gas outflow detected (\citealt{M11}) is co-spatial with the radio continuum emission (\citealt{H07}) and \textit{Chandra} X-rays (\citealt{GM09}).

\indent Our current study is focused on the central kiloparsec ($\sim$\,12\,arcsec) of NGC\,4438, at sub-arcsecond resolution across the spectrum. It aims to unveil how the central features observed at different wavelengths around the AGN are connected, thus getting insights of how feedback works. In particular, we are interested in understanding the eventual connection between the ionised gas outflows and radio jets. We propose a multi-wavelength approach considering optical imaging and spectroscopy, radio data and X-ray observations. This methodology requires high-resolution and sensitive data at the different wavelengths, not only for the morphological comparison but also for the energetic analysis. Achieving sub-arcsecond spectral index ($\alpha$) maps enables us to generate high-resolution models (see Sect. \ref{Analysis}). This makes it possible to analyse the substructures ($\sim$\,25 pc scale) of the central kiloparsec of the galaxy individually for the first time.

This paper is structured as follows. Section \ref{Data} provides the details of the data analysed in the paper. In Section \ref{Data_reduction} and \ref{Analysis}, we give a comprehensive explanation about the data reduction and the basis of the theoretical analysis used to reach our results. In Section \ref{Results}, we show the main results of the radio data analysis: flux images, spectral index maps and minimum energy parameters maps. We discuss the impact of radio jets on the host galaxy and their capability to drive ionised gas outflows in Section \ref{diss}. In the same section, we present a possible multi-wavelength interpretation to explain the feedback from the galaxy. Finally, Section \ref{conclusions} summarises the main conclusions.

\begin{table*}
\centering
\caption{Observing log of the data analysed in this paper. } 
\begin{tabular}{c c c c c c}  

\hline\hline  

 Telescope   &  Project  &  Observation date  & Frequency (Band) &  Time & Calibrators\\ 
  &   &   &(GHz) &  (s) &  \\ 

(1) & (2) &  (3) & (4) & (5) & (6) \\ 
\hline\hline
   VLA & AH230 & 24/05/1986 & 1.44 - 1.54 (L) & 2640 & 1252+119 (3C48, 3C138, 3C286) \\
   VLA  & 22A-180 & 17/06/2022   & 3.97 - 8.00  (C) & 3960  & \begin{tabular}{c}1254+1141 (3C286)\end{tabular}   \\
   VLA  & 24B-323* & 16/12/2024   & 7.97 - 12.02  (X) & 678 & \begin{tabular}{c}J1239+0730 (3C147)\end{tabular}   \\
   e-MERLIN  & CY16021* & 19-20/08/2023 & 1.25-1.75 (L) & 58223 & \begin{tabular}{c}J1230+1223\\(1407+2827, 0319+4130, 
   1331+3030)\end{tabular}  \\
\hline       
\end{tabular} 
\tablefoot{(1) Telescope and configuration, (2) project code, (3) observation date, (4) range of frequency of the observation, (5) observation time on source and (6) calibrators; the first one is the phase calibrator, flux/bandpass calibrators appear in parenthesis. \cite{H07} published AH230 data. \\
* is indicative of proprietary data.}
\label{table:1}  

\end{table*}

\section{Data}\label{Data}

For the multi-wavelength approach, we imaged and analysed high-resolution radio data from e-MERLIN\footnote{e-MERLIN: \url{https://www.e-merlin.ac.uk}} (L-band) and VLA (L, C and X-band), as well as X-ray data from \textit{Chandra}, between 0.5 and 10 keV. See Table \ref{table:1}.

\subsection{e-MERLIN data} \label{Data1}

NGC\,4438 was observed with e-MERLIN (project ID:\,CY16021, PI: Puig-Subirà). The observation consists of two epochs on 2022 August 19 and 20 with a duration of 12.95 and 15.16 hours, respectively. The Lovell telescope was not included in this run, so the number of antennas in operation was 6, providing baselines from 11 km to 217 km. It is an L-band observation centred at frequency of 1.5\,GHz with a bandwidth of 512\,MHz. The total bandwidth was divided into eight spectral windows each containing 128 channels. It was scheduled as a phase-referenced observation, where on-source scans lasted 7 min and were interleaved with 3 minute scans of the phase calibrator, J1230+1223 ( J2000 RA: 12:30:49.42 DEC: +12.23.28.04, separation 0.97$^{\circ}$, flux peak ${\sim}$ 2.8 Jy). The other calibrators involved in the observation were 1331+3030 (flux calibrator), 0319+4130 (check source) and 1407+2827 (bandpass calibrator). By blending these data with L-band data from VLA, we can create an image that takes into account the compact (e-MERLIN) and the extended (VLA) emission (see Table \ref{table:2}). 

\subsection{VLA data}\label{Data2}

For this study, we use data from the VLA archive. We have reduced and analysed three data sets involving L, C and X-band data, whose main parameters can be found in Table \ref{table:1}. In this section we describe chronologically each set. We pursue VLA data at maximum resolution, so A-configuration\footnote{VLA configuration and resolutions:\ https://science.nrao.edu/facilities/vla/docs/manuals/oss/performance} at L, C and X bands is the most appropriate option.

In  1986, before the VLA expansion, NGC\,4438 was observed in the L-band (1.44 - 1.54 GHz) within the project AH230 (PI: E. Hummel), and later analysed in \cite{H07}. The bandwidth is 100\,MHz, divided into two 50\,MHz spectral windows (1.44-1.54\,GHz). The on-source observations consist of 9 scans of 24 minutes each one. 
J1252+119 is the phase calibrator (B1950 RA: 12:52:07.71 DEC:+11.57.20.82, separation 6.7$^{\circ}$, flux peak ${\sim}$ 0.98 Jy), 3C48 and 3C138 are the flux calibrators and 3C286 is the bandpass calibrator. We re-reduced this data set using them to complement the higher resolution data (e-MERLIN proprietary data), with the objective of adding extended structures. The project AH230 also provides C-band observations not included in this work, since the current data analysed (22A-180, see details below) represent a significant improvement in terms of sensitivity, resolution and bandwidth.\\

As part of the 22A-180 project (PI: Jiangtao Li), C-band (4-8\,GHz) observations were carried out on 2022 June 17, using the VLA-A configuration. The total bandwidth is covered by 32 spectral windows of 128\,MHz. In total, 2.30 hours of phase-referenced observations were performed, including 1.10 hours on NGC\,4438. It consists of eight scans of 9 minutes each interleaved with observations of J1254+1141, the phase calibrator (J2000 RA: 12:54:38.25  DEC: 11:41:05.87, separation 6.7$^{\circ}$, flux peak ${\sim}$ 0.8 Jy beam$^{-1}$). 3C286 was used both as flux and bandpass calibrator.

We also use X-band proprietary data observed in December 2024 under VLA-A configuration project 24B-323 (PI: Puig-Subirà, M). The observation consists of 2 scans of 339 seconds on source, using J1239+0730 as phase calibrator (J2000 RA: 12:39:24.58 DEC: +07.30.17.18, separation 6.2$^{\circ}$, flux peak ${\sim}$\,0.4\,Jy\,beam$^{-1}$) and 3C147 as flux/bandpass calibrators. The aggregate bandwidth is 4.05\,GHz, (from 7.98 to 12.03\,GHz), split in 32 spectral windows of 128\,MHz each. \\

\subsection{\textit{Chandra} data}\label{Data3}

NGC\,4438 was first observed by \textit{Chandra} (P.I. Christine Jones, ObsID 2883, date: 2002-01-29) in a single exposure of 25.4 ksec (net exposure after flares removal is 25.1 ksec) to study the effect of the environment in the galaxies NGC\,4435 and NGC\,4438 within the Virgo Cluster \citep{M04}. It was then observed four times (P.I. Jiang-Tao Li, ObsIDs 21376, 23189, 23037, and 23200, between 2020-03-20 and 2020-03-28) with a total exposure time of 96 ksec (net exposure of 94.5 ksec) to study the X-ray nuclear bubble \citep{Li22}. Although it also appears in the snapshot observation with ObsID 4960 (5 ksec), the above five observations are the only ones with enough counts ($>\,$500 counts) to be used in this analysis.

\section{Data reduction}\label{Data_reduction}

In the following sections, we provide a comprehensive explanation of the reduction and analysis processes. We provide details of the different steps to ensure reproducibility of the data processing.\\\\
We first calibrated the radio data by the pipelines of the instruments, wherever possible (see next subsections). Then we self-calibrated each dataset and imaged them using CASA. For X-ray data, we used CXC \textit{Chandra} Interactive Analysis of Observations (CIAO) software\footnote{\url{https://cxc.cfa.harvard.edu/ciao}} and the calibration database CALDB to calibrate, reprocess and image the data.

\subsection{e-MERLIN data (L-band)} \label{Data_reduction1}

We calibrated the data with the e-MERLIN CASA pipeline version: v1.1.08 \citep{M21} using CASA version 5.5.0.  We processed each epoch individually. The first automatic flagging was in strict mode, so two spectral windows and one antenna were completely flagged. To avoid this, we dug into the pipeline modifying the flags file, some parameters ({\textit{`applymode'}} from {\textit{`calflagstrict'}} to {\textit{`calflag'}}, {\textit{minbl}}\,=\,2 and {\textit{minsr}}\,=\,1) and also the averaging times. For amplitude and phase calibrations we reduced the solution interval from 32 seconds to 8 seconds and for the delay calibration (K1) from 180 seconds to 32 seconds. Throughout this process, the calibration was improved, and we recovered previously flagged data. Once a satisfactory initial calibration (i.e the pipeline data products) was achieved, we concatenated the two measurement sets (see Sect. \ref{Data1}) and started the self-calibration.

The phase calibrator (J1230+1223) is slightly resolved source at 2.89 Jy. In order to recover the extended emission in the model, we first self-calibrated the phase calibrator and then we applied the improved model to the sources. This first self-calibration was divided into four steps: two phase mode calibrations and two amplitude and phase calibrations. The interval time for each round is inf, 120\,s 60\,s, 9\,s and the S/N threshold we imposed was 3.

We then applied the generated tables both to the phase calibrator and to the target. With this improved model in the data, we started self-calibrating the target.
Due to the low brightness of our target, we only carry out one round of self-calibration in phase, combining all the spectral windows with an interval time solution of 420\,s. To image and generate the model we used CASA task {\textsc{tclean}} with cell size of 0.025 arcsec.

The long baseline of e-MERLIN and the capabilities of the VLA, allow us to capture both the compact and diffuse emission in NGC4438. Furthermore, note that the e-MERLIN shortest baseline ($\sim$ 10-15 km) is similar to the baseline lengths as VLA A array, thus provide deeper images with good imaging fidelity on a wider range of scales.\footnote{See other examples of e-MERLIN and VLA combination for NGC 6217 \citep{W19} and NGC 1068 \citep{M24}.} Therefore, once the e-MERLIN dataset was calibrated, we combined it with L-band VLA data (see the calibration parameters in Sect. 2.1). We used the task \textsc{concat} to combine the visibilities, weighting the data through \textsc{statwt}. Then we created the final L-band image (e-MERLIN + VLA) using \textsc{tclean}, with a cell size of 0.02.  

\begin{table}\centering
\caption{Self calibration and imaging parameters} 
\begin{tabular}{c c c}   

\hline\hline  

Project  &  Self-cal mode  & Solint \\ 
  &   & (s) \\ 

(1) & (2) &  (3)\\ 
\hline\hline
AH230 & p & 54  \\
22A-180 & p & 54  \\
 & p & 3600  \\
 & ap & 405  \\
24B-323 & p & 339 \\
 & p & 169  \\
 & ap & 679  \\
CY16021 & p & 420 \\
\hline       
\end{tabular} 
\tablefoot{(1) Project code, (2) self calibration mode in each steep, (3) interval time solution in seconds. All these self-calibration steps are performed on NGC 4438. See Sect. \ref{Data_reduction} for the complete process and details of the calibration. For the {\textsc{tclean}} imaging we use {\textit{deconvolver\,=\,`mtmfs'}}, {\textit{weighting\,=\,`briggs'}} and {\textit{robust}}\,=\,0.5. Otherwise it is specified in the respective section.}
\label{table:selfcal}  

\end{table}

\subsection{VLA data (L, C and X band)}\label{Data_reduction2}

AH230 project was observed in 1986, so we conducted the calibration manually. We follow the open NRAO tutorial\footnote{Reduction tutorial: \url{https://casaguides.nrao.edu/index.php?title=VLA_Continuum_Tutorial_3C391-CASA6.4.1 }}. First, we conveniently split the observation measurement set file and inspect it with {\textsc{plotms}} to identify the data that should be flagged. We note that spectral window 1 on antenna VA19 required flagging. After the initial flagging, we started with the calibration, which consist of 5 steps: 1) set the flux density scale ({\textsc{setjy}}) providing a flux density value for the amplitude calibrator (3C286) according to the band; 2) generate a calibration table for the gain curve of each antenna depending on zenith-angle  ({\textsc{gaincal}}); 3) carry out an initial phase calibration to average the phase variations with time in the bandpass; it moderates the effects of variations from integration-to-integration and from scan-to-scan on the bandpass calibrator; 4) derive corrections for the complex antenna gains; 5) finally, produce the amplitude gain scaling table to determine the system response to a source of known flux density, the flux calibrator (task: {\textsc{fluxscale}}).

We applied all the generated tables to the target (task:{\textsc{applycal}}) and initiated a model by a first imaging (task: {\textsc{tclean}}). For AH230 data, we performed only one phase self-calibration loop because of the sensitivity. We chose a solution interval of 54 seconds, combining the two spectral windows. The cell size used for the imaging was  0.066 arcsec.

The calibration of the 22A-180 data was performed using the standard VLA CASA pipeline, version: 6.4.1\footnote{National Radio Astronomy Observatory. “VLA CASA Pipeline.” Available on \url{https://science.nrao.edu/facilities/vla/data-processing/pipeline} }. Once calibrated, we split the target source from the calibrated data without averaging.
We imaged the data using the task {\textsc{tclean}}, selecting a cell size of 0.03 arcsec. The first model was generated in two steps to achieve a good coverage for both compact and extended emission. In other words, we generated a first high resolution image focused on compact structures ({\textit{smallscalebias}}\,=\,0.9). Then, we created a second image to add the extend emission contribution to the model. We addressed this by restarting the same model in a second {\textsc{tclean}} run and modifying some parameters in favor of the extended structures, {\textit{uvrange\,=\,`<300klambda'}}, {\textit{smallscalebias}}\,=\,-0.5 and {\textit{robust}}\,=\,2. \\
\indent The self-calibration of NGC\,4438 was carried out in three steps: two phase self-calibrations followed by an amplitude and phase calibration. The parameters in the {\textsc{tclean}} were the same for the three images: {\textit{cellsize}}\,=\,0.02, {\textit{scale}}\,=\,[0,12,30]. In the phase self-calibrations, in order to obtain a good S/N, we combined all spectral windows in the first step and groups of 4 spectral windows in the second one. In the amplitude and phase self-calibration each spectral window was treated individually.

For the X-band data from 24B-323, we ran the standard CASA pipeline. The self calibration consisted of three steps, two rounds of phase calibration and one of amplitude and phase. In the first phase self-calibration and the amplitude and phase self-calibration step, each spectral window was treated individually while in the second phase self-calibration we combined all the spectral windows to improve the S/N. We combined the two scans in the amplitude and phase self-calibration step.

\subsection{\textit{Chandra} data}\label{Data_reduction3}

We use HEASARC archive\footnote{\url{https://heasarc.gsfc.nasa.gov}} to download ObsIDs 2883, 21376, 23189, 23037, and 23200. We then processed these observations using CXC \textit{Chandra} Interactive Analysis of Observations (CIAO) software\footnote{\url{https://cxc.cfa.harvard.edu/ciao}} in its version v4.16 (released in December 2023) and the calibration database CALDB v4.11.5 (also released in December 2023). As in \cite{Li22}, we reprocessed the raw data with the CIAO tool {\sc repro}. We first cleaned the data from background flares (i.e. periods of high background) that could affect our analysis. To clean them, we used the {\sc lc\_clean.sl}  task removes periods of anomalously low (or high) count rates from light curves from source-free background regions of the CCD. We then reprojected and combined observations to create a merged event list and exposure-corrected images. We used the image ObsID 21376 as a reference because it has the longest exposure time. We used the tool {\sc wavdetect} to search for sources and computed the transformation to align the images with sources detected within a 2 arcsec radius of the target using the tool {\sc wcs\_match} and {\sc wcs\_update}. {\sc wcs\_match} compares two sets of source lists from the same sky region and computes the transformation to align one to the other. {\sc wcs\_update} modifies the header aspect solution based on a transformation matrix. Note that the radius is set to 2 arcsec to optimise the alignment of the nuclear source.

The output of this process is an event file with all photons obtained in the five exposures. We then produced smoothed 0.5-2.0 keV images using the {\sc dmcopy} (to choose the 0.5-2.0 keV energy range) and {\sc csmooth} (to use adaptive smoothing to enhance the extended emission) tasks. This image was constructed by subdividing the default pixel of \textit{Chandra} (0.492 arcsec/pix) to reach 0.125 arcsec/pix. This is the best spatial resolution that we can achieve considering the total counts obtained (3500 counts in the 0.5-10 keV band). 

We also extracted the spectra of several regions of interest from each of the event files to investigate their emission mechanism, and for that purpose, we extracted the spectra of two circles, one for the target and another for the background in a free-source sky region using the {\sc dmextract}. We also obtained both ARF and RMF files, using {\sc mkwarf} and {\sc mkacisrmf} tools, respectively. Furthermore, we produced binned spectra of each region, with the {\sc ftgrouppha} tool, using the optimal binning scheme.

\section{Estimation of the radio emission parameters}\label{Analysis}

We use C-band VLA data (22A-180), which offer wide bands (4-8\,GHz) and high-resolution ($\sim$ 0.3 arcsec$^{2}$), to perform a reliable analysis of radio emission at $\sim$ 25 pc scales. Indeed, the location of the non-thermal radio jet and its energetic characterisation help us to describe feedback mechanisms. In this section we detail a methodology for quantifying the spectral index, $\alpha$, and modelling the radio emission through producing minimum energy parameters maps. \\

\subsection{Spectral index}\label{Spix}

The spectral index, $\alpha$, is defined as  $S_v \propto v^\alpha$  (\citealt{P70}).Its analysis allow us to identify the nature of the radio emission, separating thermal and non-thermal components.

We computed $\alpha$ across the C-band using {\textsc{astropy}} and \textsc{CASA} tasks. L and X-band data were not included in this part of the analysis, since their surface brightness sensitivity at the relevant scales is different when using matching uv distances. So, even though the maximum resolution is similar, the recovered emission is not directly comparable. Furthermore, the observations were carried out independently, with different depths and calibrators.

Firstly, we split the data set into 4 measurement sets of 8 spectral windows of 128 MHz each one. We selected the same uv-range in all of them (16.71 k$\lambda$-610.56 k$\lambda$) to ensure that we are comparing the same angular scales and we imaged each set with the same parameters. Then we smoothed higher resolution images to the lowest resolution involved (task: {\textsc{imsmooth}}). The common restoring beam for the 4 sub-images is 0.42'' x 0.36'' and the rms values are 8.3, 8.8, 11.6, 11.4  $\mu$Jy beam$^{-1}$ (from lowest to highest central frequency). The regrid step is not needed because of the homogeneous conditions in the creation of the images. Once we got comparable images, we exported them from CASA images to fits files (task: {\textsc{exportfits}}), to analyse them pixel by pixel. We imposed a S/N threshold for each pixel, such that if the ratio between the flux and the standard deviation of the residual image is $<$5, that pixel is discarded, to minimise uncertainties. Thus, pixels with a S/N $<$\,5 (3 mJy beam$^{-1}$ at 5 GHz) in any of the four images, were masked for this analysis. We calculated $\alpha$ by fitting the four values in log-log space following the relation log($S_v) \propto \alpha$ log($\nu$), using the {\textsc{NumPy}} function {\textsc{polyfit}}. The spectral index values per region are presented in Table \ref{table:3}. They were calculated by averaging the spectral index values (per pixel) within a restoring beam area. The errors account for the dispersion of the fit and the uncertainty at each point. 
We obtain compatible results when using directly CASA tools (i.e  {\textsc{calckmask}} on {\textsc{imstat}} tool with mode='spix').

\begin{table*}[!b]
\centering
\caption{Parameters of the continuum images in each band.}    
\begin{tabular}{c c c c c c c c c c c c}     
\hline\hline  
Band & Project & Central Freq. &  \multicolumn{3}{c}{Beam size} & rms & S$_{peak}$ & S$_{tot}$ \\ 
  & & (GHz)  & \multicolumn{1}{c}{Major} & \multicolumn{1}{c}{Minor} & \multicolumn{1}{c}{PA} & & & \\ 
 &  & & \multicolumn{1}{c}{(arcsec)} & \multicolumn{1}{c}{(arcsec)} & \multicolumn{1}{c}{(º)} & (mJy beam$^{-1}$) &  (mJy beam$^{-1}$) & (mJy) \\ 

(1) & (2) &  (3) & (4) & (5) & (6)  & (7) & (8) \\ 
\hline\hline 
   L & CY16021 and AH230 & 1.5 & 0.21 & 0.12 & 31.01 & 0.015 & 0.363  &  59.4 $\pm\,$ 1.78 \\

   C & 22A-180 & 6 & 0.33 & 0.30 & 8.58 & 0.004 & 0.565 & 21.69 $\pm\,$ 0.65\\
   X & 24B-323 & 10 & 0.32  & 0.21 & -55.44 & 0.009 & 0.459 & 12.15 $\pm\,  $0.36\\

\hline                  
\end{tabular}
\tablefoot{(1) Band, (2) project code of data used, (3) central frequency, (4)-(5)-(6) restoring beam parameters; major axis, minor axis and position angle, respectively, (7) rms noise, (8) peak brightness and (9) flux density, including NW-lobe and nucleus.\\
The L-band image presented in this work is the result of combining data from projects AH230 and CY16021. See the text for details.\\
}  
\label{table:2}  
\end{table*}

\subsection{Physical model of the radio emission}\label{Model}

We use a parametrisation that models the power (P) and energy (U) of the jet, the cosmic-ray energy density, $u_{min}^{CR}$, the cosmic ray electron diffusion length, $L_D$, the magnetic field strength, $B_{min}$, and the particle lifetime, $t_{syn}$. Our methodology follows an analogous approach to \cite{I99} applied to NGC\, 4438 at 1.5 arcsecond resolution in \cite{H07}. In this work, we present the analysis at $\sim$\,0.4 arcsecond, enabling the analysis of the substructures separately, in particular enabling us to distinguish the nucleus from the additional structures. The resultant maps are computed pixel by pixel since the luminosity ($L$) 
and the spectral index ($\alpha$) both vary across the galaxy. Note that $L=4\pi d^2S_{\nu}$ requires an integration over frequency which is dependent on spectral index, $S_{\nu} \propto \nu^{\alpha}$, (\citealt{P70}), which results on 

\begin{equation}
    L = 4 \pi \, \text{d}^2 \, S_{v0} \left( \frac{v_0^{-\alpha}}{\alpha + 1} \right) \left( v_2^{\alpha + 1} - v_1^{\alpha + 1} \right)
\end{equation} 

The common spectral cut-offs are \( \nu_1 = 1 \times 10^7 \) Hz and \( \nu_2 = 1 \times 10^{11} \) Hz and our reference frequency is \( \nu_0 = 4.48 \times 10^9 \)\,Hz.\\
We use three inputs to produce these parameter maps: 1) the naturally weighted flux density \footnote{The flux density is converted from Jy beam$^{-1}$ to Jy/pixel for this analysis.} map from which the luminosity is computed, 2) the spectral index map at the appropriate resolution, and 3) a map of the line-of-sight distances, {\textit{l}}\footnote{{\textit{l}} is the line-of-sight distance through the galaxy at that pixel \citep{I99}. Following \cite{H07}, we use a geometry in which the line-of-sight depth is taken to be the measured average width of the mini double-lobed source, that is, 4.21 arcsec = 344 pc.}.

Following the minimum energy assumption, one can acquire maps of $B_{min}$ and $u_{min}^{CR}$, averaged over a line of sight. These are computed in cgs units (see \citealt{P70}; \citealt{D91}, \citealt{I99}), from:

\begin{equation} 
    B_{\text{min}} = \left[ 6\pi (1+k) c_{12} V^{-1} L \right]^\frac{2}{7}
\end{equation}

\begin{equation}
    u_{min}^{CR}=(1+k) c_{12} V^{-1} B^\frac{-3}{2} L                   
\end{equation}

where $V$ is the volume ($V$  = r$^{2}$ {\textit{l}}, being $r$ the linear distance subtended by the pixel), and $L$ is the luminosity. $c_{12}$ is one of Pacholczykys constants, assuming the common cut-offs \( \nu_1 = 1 \times 10^7 \) Hz and \( \nu_2 = 1 \times 10^{11} \) Hz (\citealt{I99}; \citealt{H07}), in CGS units ($cm^{-3/4} g^{3/4} s^{-1/2}$), which, for the convention $S_v \propto v^\alpha$ is defined as

\begin{equation}
    c_{12} =  1.06 \times 10^{12} \left( \frac{2\alpha + 2}{2\alpha +1}\right) \frac{\left[ v_1^{\frac{(1 + 2\alpha)}{2}} - v_2^{\frac{(1 + 2\alpha)}{2}} \right]}{\left( v_1^{1 + \alpha} - v_2^{1 + \alpha} \right )}       
\end{equation}
The parameter $k$ is the heavy particle to electron ratio. Its value can range from 1 to 2000 depending of the system and assumptions\footnote{The dependence between $B_{min}$ and $k$ is weak ($B_{min} \propto k^{2/7}$) and well behaved in this range. Common values of $k$ are discussed on \url{https://www.cv.nrao.edu/~sransom/web/Ch5.html} \label{NRAO}}. A value of $k$ = 40 is commonly used in the literature coinciding with the value of cosmic rays  near the Earth exhibit (\citealt{ERA16}). Although it does not describe exactly the radio galaxy behaviour, we use $k$ = 40 in order to be able to compare our results with previous studies. Note that this $k$ value affects the absolute scale of the maps but not the point-to-point variations.\\

We can assume that the relativistic electrons inside the radio lobes lose most of their energy by synchrotron emission. Thus, the $t_{syn}$ of these particles can be obtained as the synchrotron cooling timescales for cosmic-ray electrons emitting at frequency $\nu$ : 
\begin{equation}
    t_{syn} \approx c_{12}(B_{min})^\frac{-3}{2}           
\end{equation}
We have to consider this value as an upper limit because if other loss mechanisms (e.g. inverse-Compton scattering) are significant, the actual source lifetime will be shortened (\citealt{ERA16}). 

To derive the radio power either of the following two equations can be applied:

\begin{equation}
    P = \frac {U}{t_{syn}} 
\end{equation}

where $U$ is $u_{min}^{CR}$ integrated along a line of sight (so $u_{min}^{CR} \times V$, for each pixel) and $t_{syn}$ is the particle lifetime or 

\begin{equation}
    P = \left ( 1 + k \right ) L
\end{equation}

In this work we use the equation 7, where $L$ is the observed luminosity at a given point.\\

Once calculated the radio power for each pixel (see Sect. \ref{Results_maps}), we determine the value of the jet power depending on the selected region. First, we select a region and sum the total radio power contribution. Then use the relationship between the total radio power and jet power presented in \cite{M07}, equation 8 in this paper. This equation is suitable for the case of NGC\,4438 and has been used in comparable analyses (see Sect. \ref{diss_energetic}:

\begin{equation}
    log(P_{jet}) = (0.81 \pm 0.11) log(P_{radio}) + 11.9 ^{+4.5}_{-4.1}
\end{equation}

Finally, to obtain the diffusion length, we need to calculate the propagation velocity of the plasma waves. For this, we chose the Alfvén intermediate mode (see also \citealt{P70} and \citealt{I99}). 

\begin{equation}
    v_a = \frac {B_{min}}{\sqrt{\mu_0\rho}} [SI] = \frac {B_{min}}{\sqrt{4\pi\rho}}  [cgs] 
\end{equation}

\noindent where $v_a$ is an order-of-magnitude estimation assuming a constant mean ISM density of 0.18 cm$^{-3}$, commonly used in the literature \citep{I99}. Since we assume the medium is ionised, we multiply this density by the proton mass to give it the proper dimensional units. From this point, the diffusion distance ($D_L$), i.e. the distance that an electron travels before it loses its energy, can be calculated as follows:

\begin{equation}
    D_L =  v_a t_{syn}
\end{equation}

In sections \ref{Results_maps} and \ref{diss_energetic}, the maps produced are described and discussed.

\section{Results}\label{Results}

In this section, we present the C, L and X-band images (Fig. \ref{fig:1}) describing the substructures of NGC\,4438 (Fig. \ref{fig:esquema}), the C-band spectral index map (Fig. \ref{fig:spindex}),  the maps of minimum energy parameters (Fig. \ref{fig:energymaps}) previously defined and the X-rays images (Fig. \ref{xray}). The main information of the C, L and X-band images can be found in Table \ref{table:2}. We analysed the images using {\textsc{imstat}} and {\textsc{imfit}} tasks for more diffuse (lobes) and more compact (nucleus) regions, respectively.

\subsection{C - band image (VLA)}\label{Results_C}
\begin{figure}[!htbp]
    \centering

    \includegraphics[width=\linewidth]{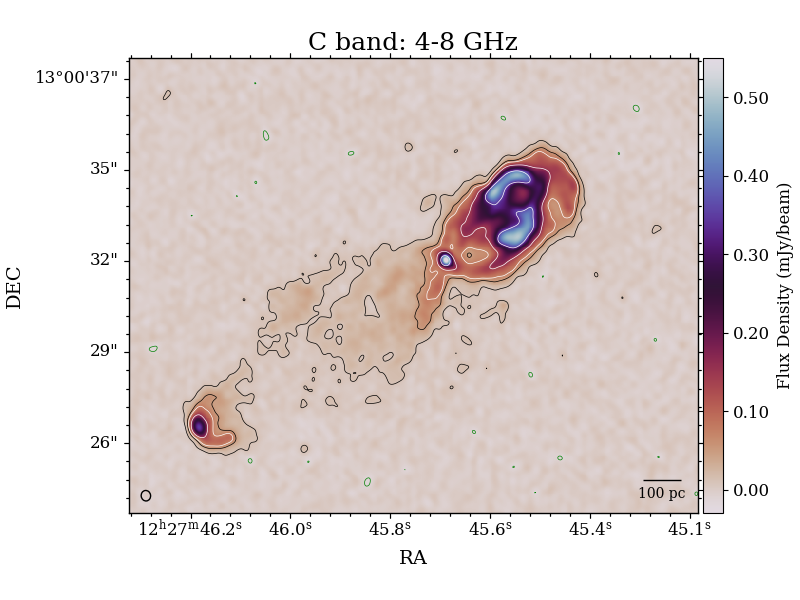}

    \includegraphics[width=\linewidth]{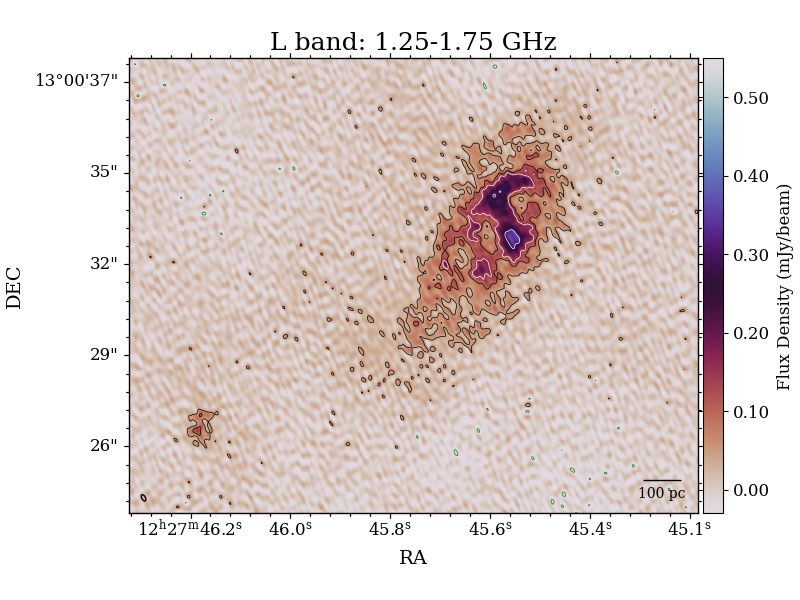}

    \includegraphics[width=\linewidth]{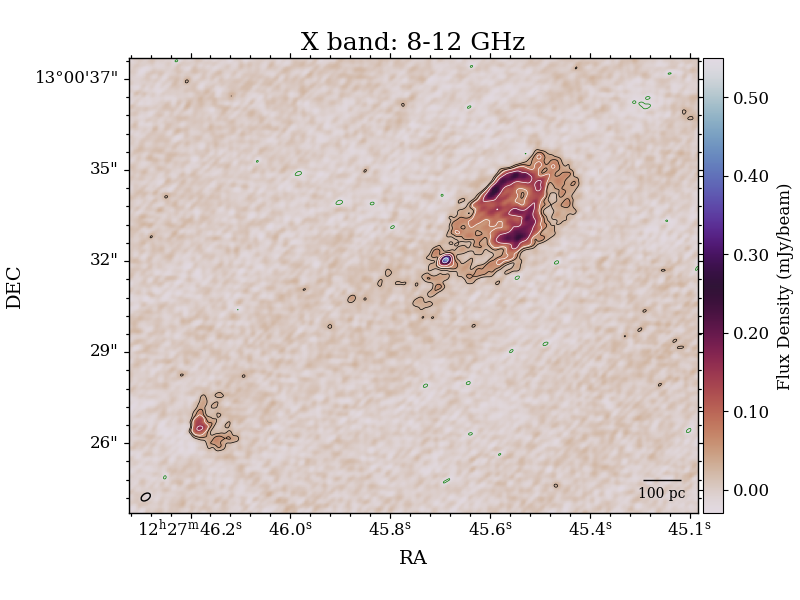}

    \vspace{1em}
    \caption{NGC 4438 radio structures. Top: C-band from VLA data. Center: L-band image from blending e-MERLIN and VLA data. Bottom: X-band image with VLA data. Contours: We represent in green -3$\sigma$ regions, the two black contours represent 3$\sigma $ and 0.05 mJy, and the three white ones 0.075 mJy, 0.15 mJy and 0.35 mJy. The ellipse in the bottom left indicates the restoring beam.  Despite the similar resolution in the C and X-band images, the diffuse emission captured in the C-band makes the image appear smoother. This is an intrinsic effect limited by wavelength.}
    \label{fig:1}
\end{figure}

\begin{figure}[!t]
    \centering
    \includegraphics[width=1.0\hsize]{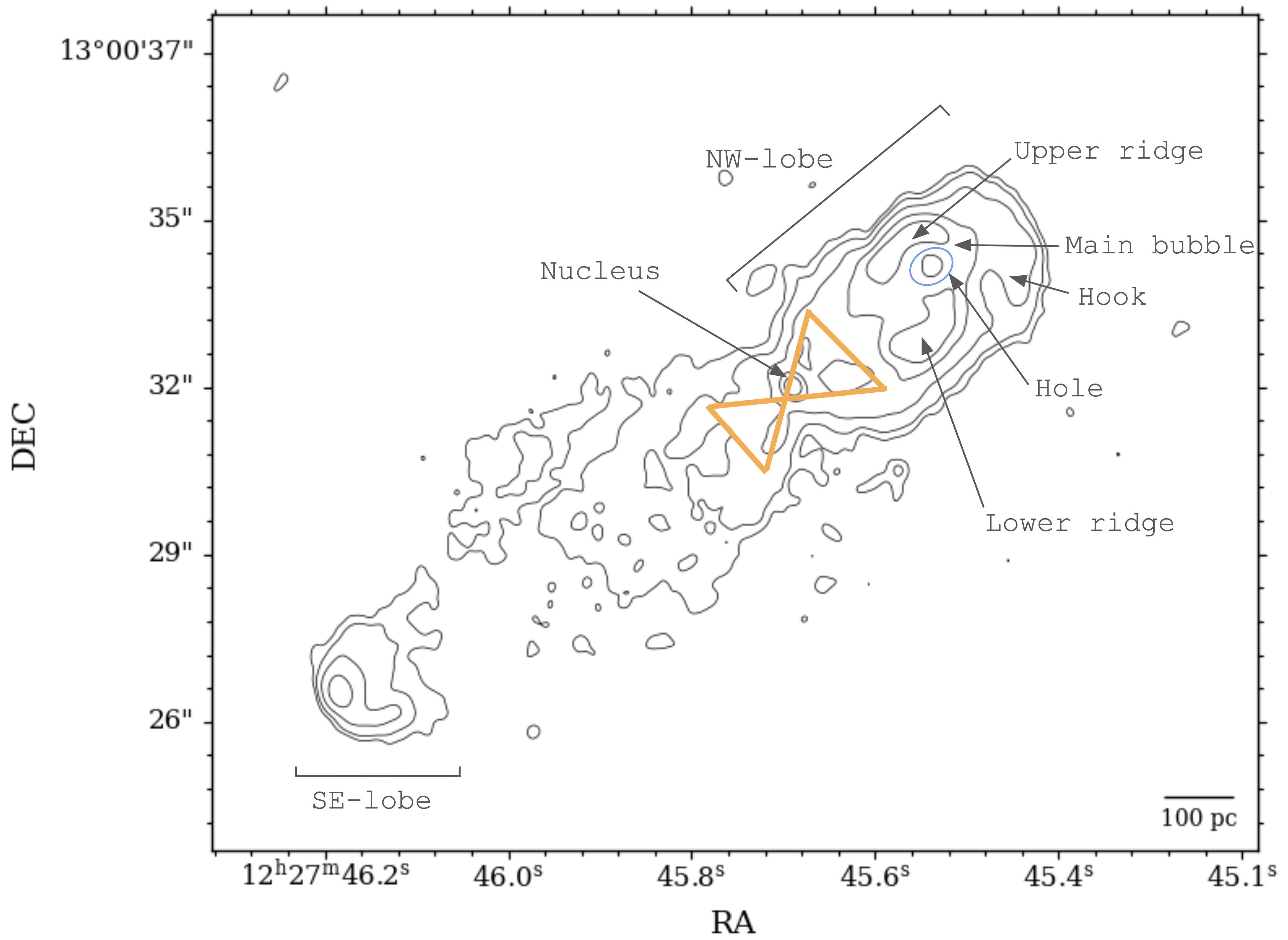}

    \caption{Scheme presenting the substructures of NGC\,4438. It is based on the C-band image (see Fig. \ref{fig:1} and Sect. \ref{Results_C}). The contours are at 3$\sigma$, 0.03 mJy, 0.08 mJy, 0.2 mJy and 0.35 mJy. The orange line shows the hourglass-like structure mentioned in the Sect. \ref{Results_C}. The NW side of the orange figure encloses the region referred to as the `base of the cone'. We name the circular cavity in the main bubble (blue circle) the `hole'.}
    \label{fig:esquema}
\end{figure}

The C-band image (Fig. \ref{fig:1}, top) obtained from the VLA-A configuration (22A-180) has a resolution of 0.33'' x 0.30'' and rms of 4 $\rm \mu$Jy beam$^{-1}$ (see Sects. \ref{Data2} and \ref{Data_reduction2}). We identify the nucleus and two asymmetric structures, namely the North-West lobe and the South-East lobe (hereafter NW-lobe and SE-lobe, respectively). We use the scheme in Fig. \ref{fig:esquema} as a reference for the different components described below.

- Nucleus: We detect a compact component at RA: 12:27:45.672 DEC: 13.00.31.680, in agreement with the C and X-band radio images quoted in \cite{H07}. It has a total nuclear radio emission of 0.91\,mJy. Extended emission arises from the nucleus in the innermost part tracing an hourglass-like structure (orange line in Fig. \ref{fig:esquema}) along the axis of the two lobes. However, the inner partial symmetry is quickly lost and the hourglass shape becomes truncated as it stretches towards each lobe, forming the bubble in the case of the NW direction.

- NW-lobe: It is the most characteristic structure because of its bubble-like shape, also visible in the HST image (Fig. \ref{RGB} and \citealt{M11}). The total extension of the NW-lobe taking into account the extended emission contribution is 4.8 arcsec ($\sim\,$ 400 pc). We distinguish three substructures within the NW-lobe, ordered from the nucleus outward: the base of the cone, the main bubble and the external hook (see Fig. \ref{fig:esquema}). According to the top panel in Fig. \ref{fig:1}, we refer as the main bubble to the bluest structure (2.4 arcsec $\sim\,$ 200 pc long). Within the main bubble, we point out the hole, a circular central cavity of 0.3 arcsec. The two most luminous structures that delimit the main bubble are defined as the upper and lower ridges. Other papers use the term `shells' \citep{H07}. The base of the cone is the fainter structure that connects the AGN to the main bubble. It rises from the nucleus to a distance of 1.8\,arcsec, until it is dominated by the lower ridge where the main bubble emerges. Finally, in the most north-western side, we can see a more extended hook-shaped structure, that represents the end of the NW-lobe. This radio image reveals a clear connection between the lobe and the nucleus, showing how the bubble emerges from it forming an hourglass-like structure. This complements the optical images, where the effect of the dust obscures this central area.

- SE-lobe: It is situated at 9.5\,arcsec ($\sim\,$ 730 pc) far from the nucleus. It is smaller and weaker than the NW-lobe. In SE-lobe, we glimpse a small arc structure, measuring a flux density of 1.55 mJy, but the majority of the flux is centred in the core of the lobe.

\subsection{L - band image (e-MERLIN + VLA)}\label{Results_L}

The L-band image showed in Fig.\ref{fig:1} (middle) is obtained as a result of a combination of VLA (AH230) and e-MERLIN (CY16021) data. The former data provides the extended emission; the resulting image can be found in in \cite{H07}. The second one allows us to detect the compact structures (see Sects. \ref{Data} and \ref{Data_reduction1}).
The L-band image has a resolution of 0.21'' x 0.12'' and a rms of 0.015 mJy beam$^{-1}$.  The substructures of the galaxy may appear less connected due to the stronger contribution of the long baseline set in the e-MERLIN array, compared to the VLA.

- NW-lobe: It is still present at lower frequencies. Regarding the main bubble, emission in the upper and lower ridges persists but exhibits different shapes compared to the C-band image. The upper ridge largely retains its structure, while the emission of the lower ridge is concentrated in a central spot, being the strongest emitting region in this band. The hole is observed in the image as a region with fainter emission. A vertical structure, which may represent the compact structure that would close the main bubble, appears connecting the two ridges. It separates the main bubble and the base of the cone. In this image, the hourglass is not well detected.
 
- Nucleus and SE-lobe: The AGN and the S-E lobe are faint in this band but still detected. The $S_{peak}$ of the nucleus is 0.133 mJy beam$^{-1}$ (8.8 $\sigma$) and the S-E lobe peak is 0.128 mJy beam$^{-1}$ (8.5 $\sigma$).

\subsection{X - band image (VLA)}\label{Results_X}

From 24B-323 project VLA-A configuration data, we generated the X-band image (see Sects. \ref{Data2} and \ref{Data_reduction2}). The resolution of this image is 0.32''x 0.21'' and the rms is 0.009 mJy beam$^{-1}$ (Fig. \ref{fig:1}, bottom). This image compared to lower frequency images from VLA, resolve out most of the diffuse emission because the longer baselines in units of wavelength filter out large spatial scales.

- Nucleus: In this image, we detect compact emission in the nuclear region, being the brightest structure in X-band with total emission of 0.57 mJy. Thanks to the new data, we show how the nucleus is connected with the NW-lobe, also at these frequencies. In fact, at X-band frequencies, \cite{H07} reported a small jet-like extension ($\sim$ 0.4\,arcsec; 33 pc) at PA of 223$^{\circ}$. We also detect this feature but since we recover more emission we see how it is part of the hourglass-like structure.

- NW-lobe: At higher frequencies, since our sensitivity to extended emission decreases, we obtain a more abrupt structure that resembles the `shell' described in the literature. The main bubble is detected and the two ridges are well defined. The base of the cone is faded but still present, connecting the nucleus and the main bubble. We can see the internal structure of the hook, that persists at high frequencies.

- SE-lobe:  We detect the lobe as the arc-like feature mentioned in C-band description.

\subsection{Spectral index map} \label{Results_sp}

\begin{figure}[!h]
    \centering
    \begin{subfigure}[t]{0.48\textwidth}
        \centering
        \includegraphics[width=\linewidth]{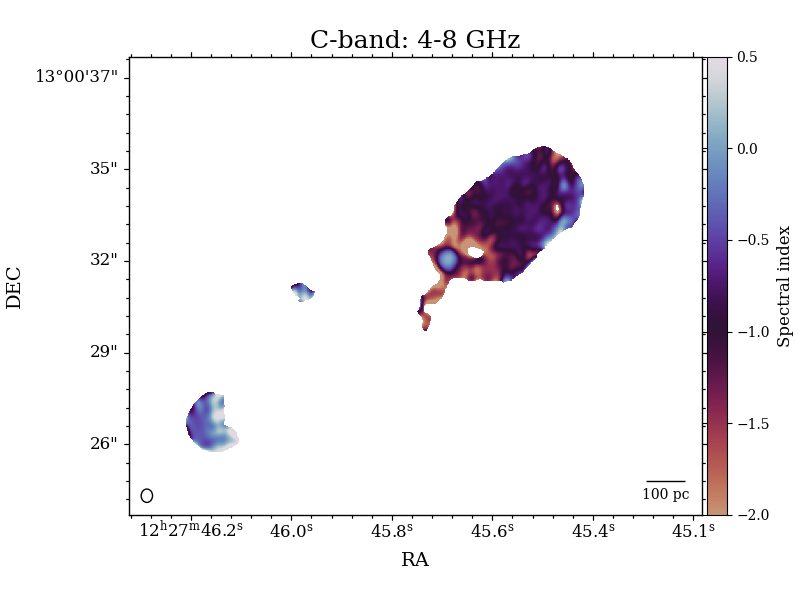}
    \end{subfigure}
    \hfill
    \begin{subfigure}[t]{0.48\textwidth}
        \centering
        \includegraphics[width=\linewidth]{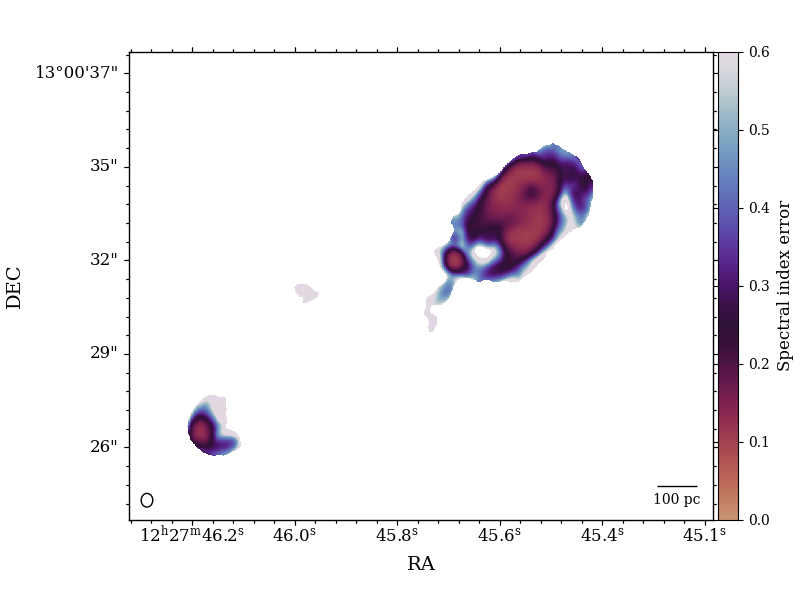}
    \end{subfigure}
    \caption{Top: C-band spectral index map created pixel by pixel with a S/N threshold of 5. Bottom: Error in the spectral index value for each pixel.}
    \label{fig:spindex}
\end{figure}

We show the C-band spectral index map (top) and its error (bottom) in  Fig.\ref{fig:spindex}. See Sect. \ref{Spix} for details of the spectral index calculations and the frequency selection. The values averaged per region can be found in Table \ref{table:3}. The map displays mostly negative values in the whole structure ranging from -2 to 0.5, with errors around 0.1, rising to 0.5 at the boundaries. Pixels with errors greater than 0.5 are not included in the discussion. 

In AGN contexts, flat or positive values ($\alpha$\,$\geq$\,-0.1) are associated with thermal or free-free emissions, whereas negative spectral indices ($\alpha$\,$\leq$\,-0.8) indicate non-thermal emission \citep{P70}. However, any spatially overlapping contribution from thermal emission would tend to flatten the observed spectral index relative to the intrinsic non-thermal value. To assess if this is happening, we estimate the flux density of radio emission associated with the thermal component (equation 6 in \citealt{C20}) from the observed H${\alpha}$ flux. We use an H${\alpha}$ flux of $\sim$\,7.68$\times$10$^{-13}$\,ergs\,cm$^{-2}$\,s$^{-1}$, which is the sum of the three H${\alpha}$ components in \cite{HM24}, covering the field of view of MEGARA (see Sect. \ref{diss_connection}). We assume the electronic temperature of this ionised gas to be 10$^{4}$ K. The calculated flux density in this region due to thermal emissions, at 5 GHz, is 0.75 mJy. By contrast, the measured flux density at this frequency, when considering only the NW-lobe region (i.e. a smaller area), is $\sim$\,23 mJy. Therefore, the inferred thermal component would represent at most 3$\%$ of the total. This free-free contribution would produce a maximum flattening of $\Delta\alpha \sim$ 0.05, below our $\sigma$ sensitivity of 0.13. Therefore, the measured $\alpha$ = -0.83 $\pm$ 0.13 is fully compatible with pure synchrotron emission.  To achieve a significant $\alpha$ flattening, beyond the error tolerance, namely $\alpha$\,=\,-0.5, the required thermal flux would be $\sim$\,9\,mJy, significantly higher than what we observe. \cite{H07} discuss this question used the method described in \cite{L01}; they calculated the electron density from the thermal flux required to flatten $\alpha$, and found it to be much higher than they expected.

The values of $\alpha$ in the NW-lobe (see Table \ref{table:3}) are consistent with non-thermal emission. We can distinguish two behaviours, consistently with the previously described structures; in the main bubble the values are between -0.7 and -0.9 (errors around 0.12) compatible with the nature of the jet ($\alpha$\,=\,-\,0.8, synchrotron emission), while in the base of the cone the values are more negative, with an average of -1.75 (with significant dispersion, i.e > 0.5, {{beyond our tolerance range for the forthcoming analysis}}). 

The AGN is clearly identified as a flat-spectrum region ($\alpha=0.07\pm 0.12$), due to the superposition of multiple self-absorbed synchrotron components, each with its own spectral turnover, distributed across the observing band (\citealt{F00}; \citealt{N02}). The nucleus is not particularly absorbed at low frequencies, with values of $\alpha$ close to 0, becoming negative as the distance from the centre increases.

The SE-lobe also shows negative values ($\alpha$ = -0.44 $\pm 0.14$), compatible with the synchrotron emission but closer to zero than the NW-lobe.

\begin{table}[!h]  
\caption{Global properties of the spectral index maps (see Fig. \ref{fig:spindex}) and C-band images (see Fig. \ref{fig:1}) for the regions identified in NGC\,4438 (see Sect. \ref{Results}).}     
       
\begin{tabular}{c c c c}   
\hline\hline       
                    
Region & $\alpha$  & S$_{peak}$ & S$_{tot}$ \\  
& (mean)   & (mJy beam$^{-1}$) &  (mJy) \\
\hline \hline                        
    NW-lobe & ... & 0.5 & 19.7 $\pm\,$ 0.59\\ 
     \hline 
    Main bubble  & -0.82 $\pm\,$ 0.13 & 0.5 & 11.1 $\pm\,$ 0.33 \\
    \hline    
    Nucleus & 0.07 $\pm\,$ 0.12 & 0.54 & 0.91 $\pm\,$ 0.07 \\
    \\
    \hline  
    SE-lobe & -0.44 $\pm\,$ 0.15 & 0.25 & 1.55 $\pm\,$ 0.05\\
    \hline                  
    \end{tabular}  
    \tablefoot{ $\alpha$ (mean) , C-band peak flux in Jy beam$^{-1}$ and C-band flux density in mJy per region. The global NW-lobe spectral index is not provided due to the high error ( > 0.5 ) when the entire structure is included.}
    \label{table:3}     

\end{table}

\begin{figure*}[!ht]
    \centering
    \makebox[\textwidth][c]{%
        \begin{subfigure}[t]{0.36\textwidth}
            \centering
            \includegraphics[width=\linewidth]{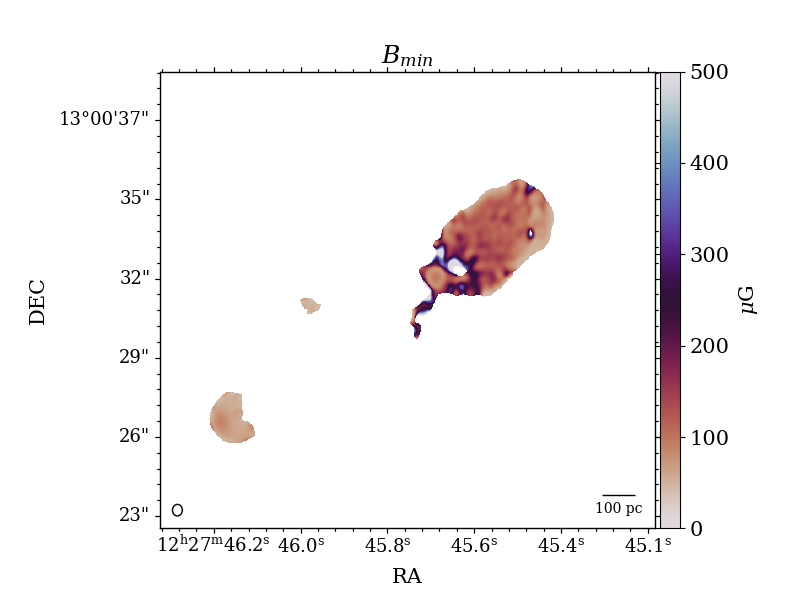}
            \caption{}       
        \end{subfigure}\hspace{0.001\textwidth}%
        
        \begin{subfigure}[t]{0.36\textwidth}
            \centering
            \includegraphics[width=\linewidth]{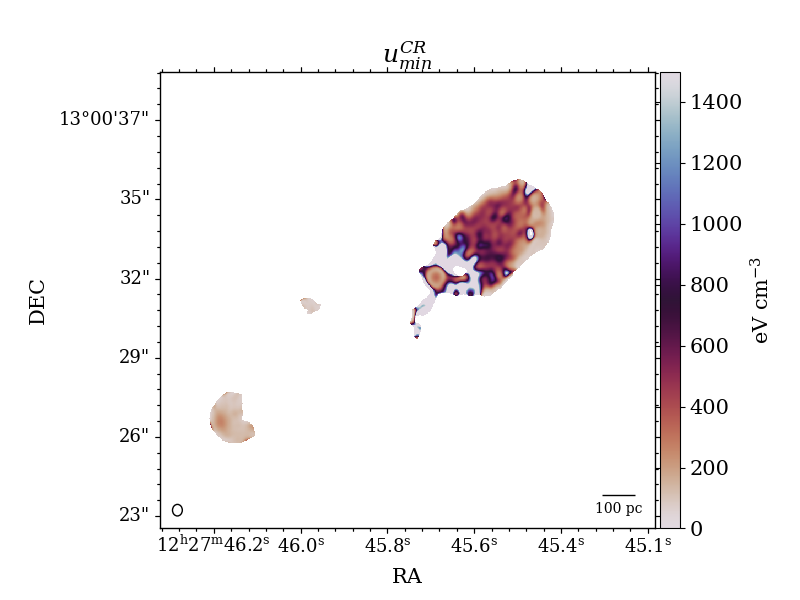}
            \caption{}
        \end{subfigure}\hspace{0.001\textwidth}
        \begin{subfigure}[t]{0.36\textwidth}
            \centering
            \includegraphics[width=\linewidth]{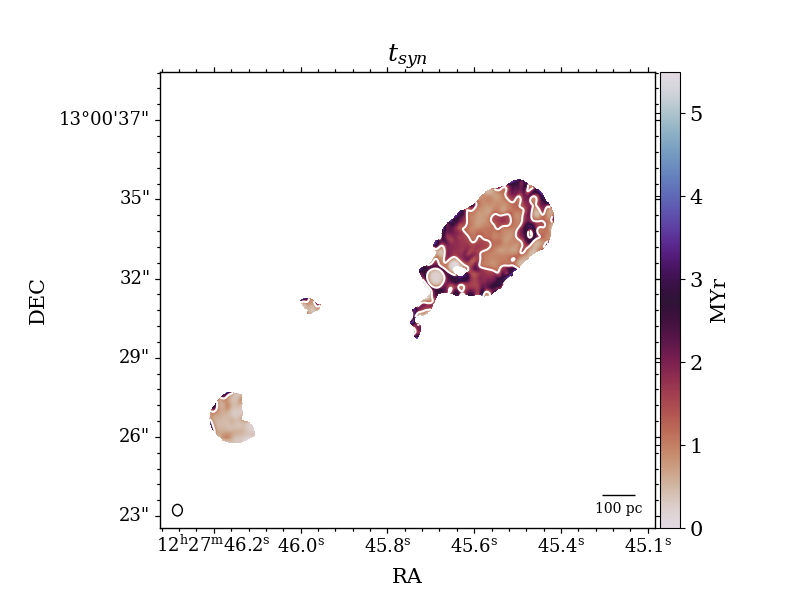}
            \caption{}
        \end{subfigure}\hspace{0.001\textwidth}%
     }

    \makebox[\textwidth][c]{%
       
        \begin{subfigure}[t]{0.36\textwidth}
            \centering
            \includegraphics[width=\linewidth]{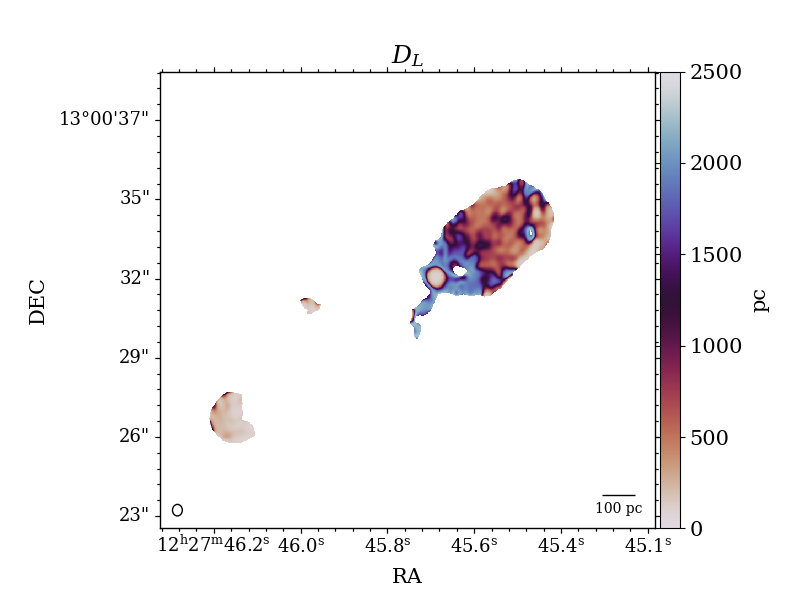}
            \caption{}
        \end{subfigure}\hspace{0.001\textwidth}
        \begin{subfigure}[t]{0.36\textwidth}
            \centering
            \includegraphics[width=\linewidth]{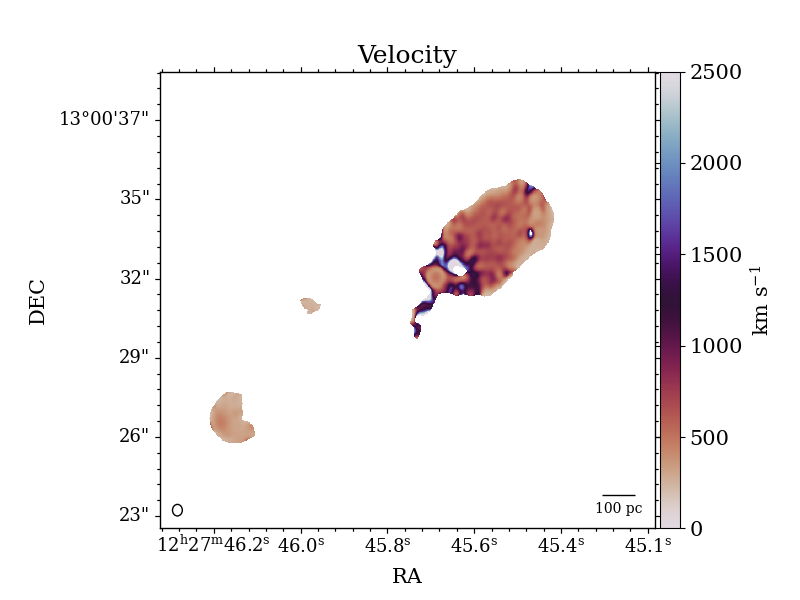}
            \caption{}
        \end{subfigure}\hspace{0.001\textwidth}
        \begin{subfigure}[t]{0.36\textwidth}
            \centering
            \includegraphics[width=\linewidth]{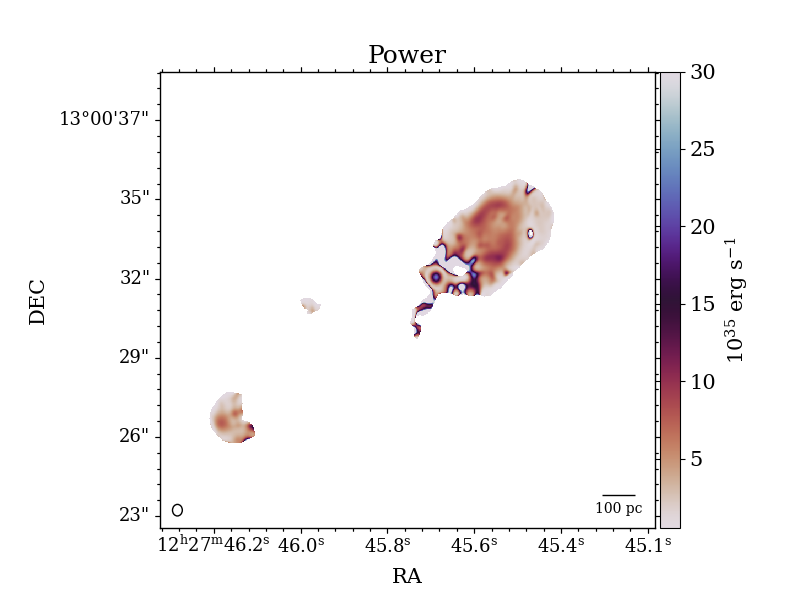}
            \caption{}
        \end{subfigure}\hspace{0.001\textwidth}
     }

    \caption{Maps of the minimum energy parameters assuming k = 40.
    (a) Magnetic field strength, $B_{min}\,$, in $\mu$G. (b) Cosmic-ray energy density, $u_{min}^{CR}\,$, in eV\,cm$^{-3}$. (c) Lifetime of the particles, $t_{syn}\,$, in Myr. The contour is 1.3 Myr. (d)  Diffusion length, $D_{L}\,$, in pc. (e) Velocity in km s$^{-1}$. (f) Cosmic-ray power, P, in erg s$^{-1}$. }
    \label{fig:energymaps}
    
\end{figure*}

\subsection{Minimum energy parameters maps}\label{Results_maps}

Following the model developed in Sect. 4.1.2, we obtain the minimum energy parameter maps (namely $B_{min}$, $u_{min}^{CR}$, t$_{syn}$, velocity, $D_L$ and power, see Eqs.1 to 9), which are presented in Fig.\ref{fig:energymaps}. In this analysis we focus in three regions, introduced in Sect. 5.1: the main bubble, the nucleus and the SE-lobe. The maximum, minimum and average values for all parameters per region are presented in Table \ref{table:4}. The connection between the AGN and the main bubble (i.e. the base of the cone) and the more extended emission (hook) are not discussed due to the large uncertainty in the spectral index, likely resulting from the low S/N in the highest-frequency sub-bands of the C-band.

In the $B_{min}$ and $u_{min}^{CR}$ maps (panels (a) and (b) in Fig. \ref{fig:energymaps}) we observe the three structures with smooth internal variations. Both the nucleus and the SE-lobe show lower values compared to the NW-lobe. The maximum of $B_{min}$ and $u_{min}^{CR}$ within the main bubble coincides with the minimal flux density, where the hole is located. When we analyse $t_{syn}$ and $D_{L}$ maps (panels (c) and (d) in Fig. \ref{fig:energymaps}) we see they are less homogeneous and present tendencies. In the $t_{syn}$ map we add a contour at 1.3 Myr to highlight the special behaviour in the main bubble, which is younger than the base of the cone. Consequently we note a co-spatial behaviour in the $D_{L}$ map, around values of 1000 pc. 

We calculate the particle velocity required to reach each position if the relativistic particles were supplied by the radio nucleus, using the positions and the  $t_{syn}$ of the particles. We choose three locations in the NW-lobe: the farthest part of the bubble (hook), the hole and the region of the main bubble closest to the AGN. The distances from the nucleus to this regions are 304 pc, 259 pc and 165 pc, respectively. Then the required velocities, calculated according to $t_{syn}$, are 330 km\,s$^{-1}$, 147 km\,s$^{-1}$ and 165 km\,s$^{-1}$. The distance to the SE-lobe is 750 pc and the minimum velocity required is 1435 km\,s$^{-1}$. These are lower limits, as we are not correcting the velocity component from possible projection effects. In all cases, v $<<$ c, so in situ acceleration for dragging the gas to their positions is not required. If we compare these calculations with the velocity map (panel (e), Fig. \ref{fig:energymaps}) computed assuming the Alfvén mode approximation, the values in the NW-lobe are in the range 600 - 700 km\,s$^{-1}$, but in the SE they are lower, around 300 - 400 km\,s$^{-1}$. That is consistent with electrons losing velocity along their travelling path. 
\begin{table}[!h]  
\centering
\caption{Minimum energy parameters per region (see Sects. \ref{Analysis} and \ref{Results}, and Fig. \ref{fig:energymaps}).} 
\begin{tabular}{c c r r r}     
\hline\hline       
Parameter& Region & Mean& Min & Max\\  
\hline           
\hline   

$B_{min}\,$  & Main bubble & 121 & 82 & 160 \\  
($\mu$G) &Nucleus &  82 & 70 & 99\\  
&SE-lobe & 72 & 56 & 89 \\  
\hline

$u_{min}^{CR}\,$ & Main bubble & 494  & 221 & 872\\  
(eV\,cm$^{-3}$)&Nucleus & 224 & 168 &  325 \\  
&SE-lobe & 176 & 106 & 266  \\  
\hline

$t_{syn}\,$& Main bubble & 1.07 & 0.71 & 2.06 \\  
 (Myr)& Nucleus & 0.47 & 0.20 & 1.31 \\  
& SE-lobe & 0.52  & 0.38 & 0.81\\ 
\hline
Velocity & Main bubble & 624 & 421 &  835 \\  
(km\,s$^{-1}$) &Nucleus & 430   & 360 & 510 \\  
&SE-lobe & 373 & 291 & 461 \\ 
\hline                  

$D_{L}$& Main bubble & 692 & 321 & 1370* \\  
 (pc) &Nucleus & 196  & 104 & 582 \\  
&SE-lobe & 200 & 121 & 299 \\
\hline

P & Main bubble & 6.59 & 3.21 & 11 \\  
($10^{35}$ erg\,s$^{-1}$)& Nucleus & 9.05 & 2.23 & 22 \\  
& SE-lobe & 4.77 & 2.76 & 7.6 \\ 
\hline  \hline  
Total Power & Main bubble &  68.1  &  \\ 
($10^{38}$ erg\,s$^{-1}$) & Nucleus & 6.7 &  \\ 
 & SE-lobe & 6.6 &  \\ 
    \hline                  
    \end{tabular}
    \tablefoot{* this value is reached at the hole location.}
    \label{table:4}
\end{table}

The power of the radio emission (panel (f) in Fig. \ref{fig:energymaps}) shows similar features as the flux. However, they do not match exactly, as the luminosity computation requires an integration over frequency that depends on the spectral index. The highest values are found in the main bubble, with  6.8 $\times 10^{39}$ erg\,s$^{-1}$, while the nucleus and the SE-lobe have integrated values of 6.7 and 6.6 $\times 10^{38}$ erg\,s$^{-1}$, respectively. The total radio power accounting for the three structures is 8.14  $\times 10^{39}$ erg\,s$^{-1}$, which is a lower limit to the total power of the source because the extended emission between the nucleus and the SE-lobe is not included due to the S/N cut-off. Furthermore, within the NW-lobe, only the main bubble is considered, excluding the base of the cone and the outermost regions. We calculate the kinetic (jet) power using Eq. 8. The case of high resolution data for this analysis, allow us to distinguish the kinetic power of each substructure. For the main bubble, it is $\sim\,1.46\times 10^{44}$\,erg\,s$^{-1}$. See discussion in Sect.\ref{diss_energetic}.

We have calculated the brightness temperature ($T_b$) following \cite{F00}. We use the peak flux density and resolution for both C and X-band images, where the maximum emission is in the nucleus. We get $T_b$ of $1.8 \times 10^2$\,K and $9.1 \times 10^1$\,K in C and X-band, respectively. These values are several orders lower than $10^5$ K, which is a common threshold used to identify black hole accretion \citep{C91}. Our results of resolution $\sim 0.3\,$arcsec do not pose a constraint on the nature of the core, since the scope of this criteria would require microarcsecond resolution for such luminosities.

\begin{table*}[!t]  
\centering
\caption{Parameters of X-ray spectral fitting  (see Fig. \ref{xray}).} 
\begin{tabular}{c c c c c c}  

\hline\hline      

Region & model  & N$_{H}$  & kT & $\Gamma$ & $\chi^2/d.o.f$\\  
&    &  (10$^{22}$ cm$^{-2}$)  & (keV) & &  \\
\hline \hline                        

Nuclear (1) & MEKAL + PL  & 0.7 $\pm\,$ 0.1 $/$ > 1.8 & 0.67 $\pm\,$ 0.08 &  2 & 9.53/12 \\ 

     \hline 
    North (2)  & MEKAL & 0.5 $\pm\,$ 0.2 & 0.7 $\pm\,$0.2 & ... & 17.1/9 \\
    & MEKAL + PL &  $<$ 0.4 & $<$ 1 & 2 &  13/8\\
    \hline  
   
    West (3) & MEKAL & 0.4 $\pm\,$ 0.2 & 0.6 $\pm\,$ 0.2 & ... & 9.1/7 \\
    & MEKAL + PL &  $<$ 0.4 & 0.8 $\pm\,$0.1 & & 7/6\\

    \hline  
    Northwest (4) &  MEKAL & 0.4 $\pm\,$ 0.2 & 0.6 $\pm\,$ 0.2  & & 9.1/7\\
     & MEKAL + PL &  $<$ 0.4 & 0.8 $\pm\,$ 0.1  & 2 & 7/6\\

    \hline              
    External (5) & PL & ... & ...  & 2.2 $\pm\,$ 0.3 & 15.75/13\\

    \hline   
    \end{tabular}  
    \tablefoot{(1) Region, (2) model, (3) foreground absorption column density (N$_{H}$), (4) temperature (kT), (5) photon index ($\Gamma$), (6) $\chi^2$/degree of freedom (d.o.f.) }
    \label{table:xrays}     

\end{table*}

\subsection{X-ray results}\label{Results_xray}

\begin{figure}[!h]
       \centering
        \begin{subfigure}[t]{0.48\textwidth}
            \centering
            \includegraphics[width=\linewidth]{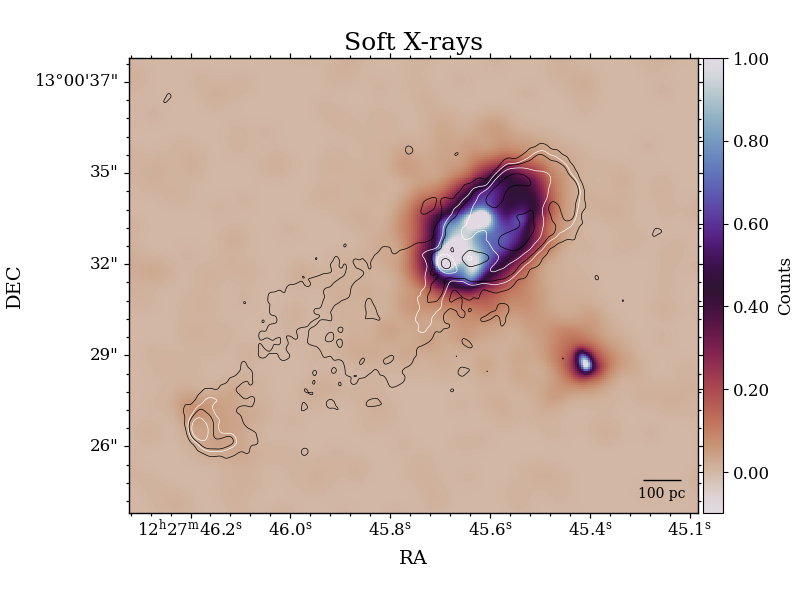}
        \end{subfigure}%
        \hfill
        \begin{subfigure}[t]{0.48\textwidth}
            \centering
            \includegraphics[width=\linewidth]{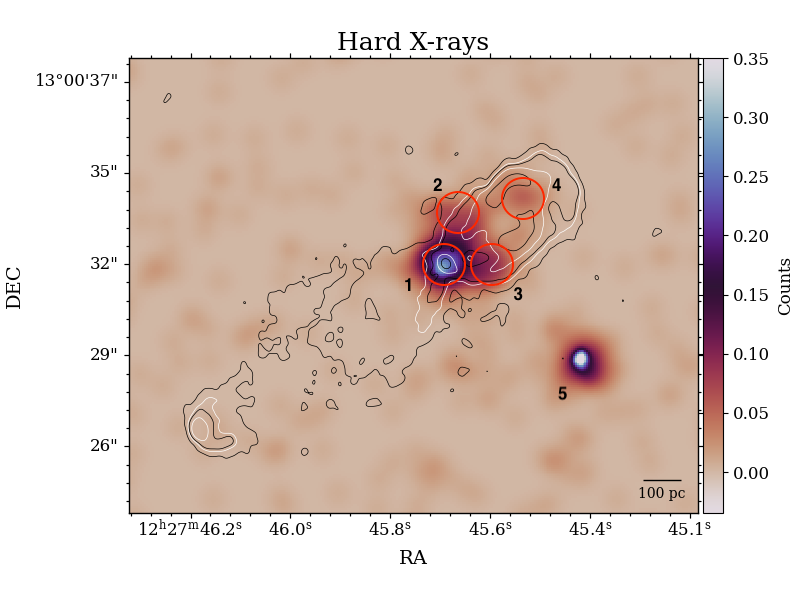}       
        \end{subfigure}
        \caption{X-ray \textit{Chandra} images with radio C-band contours. Top: Soft X-rays (0.5-2 keV). Bottom: Hard X-rays (2-10 keV) (see Sects. \ref{Data_reduction3} and \ref{Results_xray} for details). Contours: 3$\sigma $ (black), 0.05 mJy (white), 0.075 mJy (black), 0.15 mJy (white) and 0.35 mJy (black). The 0.35 mJy contour locates the nucleus.
        In the bottom panel, we indicate the five regions in which we have fitted spectra: (1) Nuclear, (2) North, (3) South, (4) West and (5) External. \\ }
        \label{xray}
\end{figure}

The soft X-ray image (Fig.\ref{xray}, top) reveals an extended morphology ($\sim$ 4.5 arcsec, i.e, $\sim$ 375 pc) that is co-spatial with the NW radio lobe and the outflow traced by H${\alpha}$ emission. The emission reaches the maximum values at the nucleus and the base of the cone, and decreases outwards. The SE-lobe can be appreciated, albeit extremely faintly.\\
\indent In the hard X-ray image  (Fig.\ref{xray}, bottom), point-like emission is detected in the nucleus, providing direct evidence of the presence of an AGN (\citealt{GM09}). We expected to detect only nuclear compact emission, considered as hard emission, but instead we also observe fainter extended emission at the base of the cone and at the hole of the main bubble. \\

We analyse both soft and hard emission, by fitting their spectra that enable us to interpret the nature of X-ray emission. We extract five regions (see Fig. \ref{xray}), including the nucleus (region 1), the brightest regions at the hard band within the NW-lobe (regions 2, 3, and 4), and an unconnected point-like source toward the south-west of the nuclear region (region 5, RA:12:27:45.376 DEC:13.00.28.599). Regions 2 and 3 (north and west, respectively) are selected at the base of the cone, while region 4 is a bright X-ray spot corresponding to a hole in the radio map (see the discussion). For spectral modelling, we use three models. Regions where a power-law (PL) is needed are considered true hard emitter regions with a non-thermal origin. Otherwise, regions where a single MEKAL model is enough to properly adjust the spectra are consistent with thermal emission. We also combine MEKAL and PL to consider the contribution of both thermal and non-thermal emission. The main results of the spectral fit are included in Table \ref{table:xrays}. PL results are not provided because they give poor fits to all the spectra. Although the MEKAL+PL seem to improve the final fit compared to thermal, its improvement compared to MEKAL model is not significant for any of the sources, except for the nucleus\footnote{Note that neither MEKAL nor PL gave satisfactory fits for the nucleus.}. Therefore, non-thermal emission is confined only to the nucleus (region 1). Although this analysis is not fully consistent with that published in \cite{Li22}, it is important to remark that they assumed a MEKAL+PL and that our reported values for the temperature are consistent with theirs when MEKAL+PL is assumed, although the regions selected are different. While \cite{Li22} took slides from the centre towards the outer parts, we have selected specific regions for this work, some of which were on the same slide and may therefore be blurred. Furthermore, we did not adjust the two models (MEKAL+PL) by default in this study; we only added the PL model if it was required for the fit. In any case, the final results are consistent, taking into account the difference in methodology.

This selection allows us to study the base of the cone in particular. The small jet-like extension that we mentioned in Sect. \ref{Results_X} and we show how is connected with the NW-lobe, commit to the major clues to precents a jet according to \citep{K10}: flatter radio and harder X-ray spectra.

\indent As mentioned above, there is also emission detected at RA:12:27:45.376 DEC:13.00.28.599, this is $\sim$ 5.1\,arcsec ($\sim$ 420 pc) from the nucleus in the SW direction (external region, 5). This feature does not align with the studied morphology of the galaxy and is not detected in radio wavelengths nor H${\alpha}$ images. We analysed the spectrum of this source, which is consistent with an absorbed power law. Assuming the same distance as that of NGC\,4438, we obtain that the X-ray 2-10 keV luminosity of the source is $\sim$ 10$^{36}$\,ergs s$^{-1}$ (photon index $\Gamma$\,=\,2.1$\,\pm\,$0.3 and N$_{H}$ < 0.6 10$^{22}$ cm$^{-2}$). This luminosity is consistent with being an X-ray binary system. Note that to be compatible with an ULX (i.e. 10$^{39}$\,ergs s$^{-1}$), the source would have to be at z > 0.004 (inconsistent with being part of NGC\,4438). A deep analysis of this source would be necessary to disentangle the origin of this emission, but this is beyond the scope of this work.

\section{Discussion}\label{diss}

\begin{figure}[h]
   \centering
    \includegraphics[width=1\hsize]{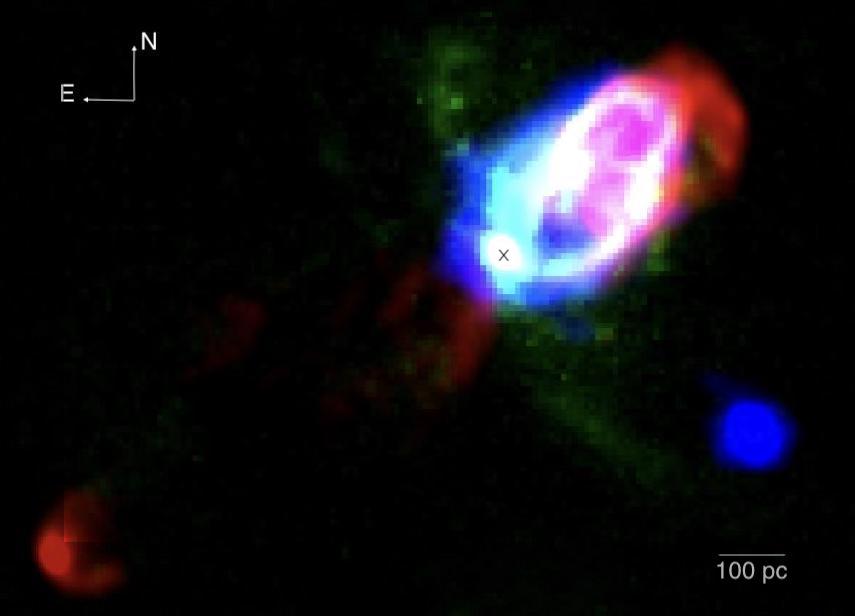}
      \caption{Composite RGB image of NGC 4438 using H$\alpha$ emission (green), C-band radio emission (red), and X-rays in the 0.5-10 keV energy band (blue). The optical data were taken from the HST archive. The C-band image is the one presented in this paper. The X-ray image is the sum of the two components (hard and soft) shown in Fig. \ref{xray}. }
         \label{RGB}
\end{figure}

Fig.\ref{RGB} shows a composite RGB image that combines the C-band radio image presented in this work, 0.5-10 keV X-ray \textit{Chandra} data, and the H$\alpha$ image from HST, all with comparable spatial resolutions. This co-spatial emission motivates us to look for physical connections between the different frequencies. For this purpose, we complement the morphological analysis with H$\alpha$ kinematic information (see Fig.\ref{components}) from the IFS study with MEGARA/GTC data published in \cite{HM24}. In this section, we discuss and propose a physical interpretation of our results through a multi-frequency comparison in terms of morphology and energetics.

\subsection{Morphology}\label{diss_morph}

In the general framework of AGN, jetted structures are prevalent in LINERs compared to brighter LLAGN, as pointed out by \cite{LEMMINGSI}. The spectral index analysis in Sect.\,\ref{Results_sp} shows a flat spectral index in the nucleus, and steep spectral indices consistent with synchrotron emission in the lobes, compatible with a bipolar jet originating from the AGN in NGC\,4438. The double-lobe morphology on the kiloparsec-scale of NGC\,4438 is detected at different wavelengths (\citealt{K02}; \citealt{H07} or \citealt{Li22}).\\

The SE radio lobe appears to propagate to larger distances, suggesting that it is able to escape without resistance through a lower-density channel. Thus, the asymmetry between the two radio lobes is hence most likely due to an asymmetric distribution of interstellar matter in the nuclear region \citep{HS91}. The non-detection of X-rays and H$\alpha$ in the SE-lobe, as shown in Fig. \ref{RGB}, supports this hypothesis. We interpret the SE-lobe as a receding lobe that appears fainter because of the absorbing material present in this region. Indeed, in HST optical images, we can detect a large amount of dust in the central part of the galaxy. The flatter $\alpha$ in the SE-lobe compared to the NW-lobe is still consistent with non-thermal emission. It can be related to younger synchrotron emission (consistent with the relative $t_{syn}\,$, Fig. \ref{fig:energymaps}), local re-acceleration of relativistic electrons \citep{CA91} or low-frequency absorption.\\

Although the nature of the emission strongly suggests that a potential jet may trigger this morphology, it is not collimated. \citet{HS06} defines radio bubbles as structures whose width is larger than approximately half the size of their length. In this way, they distinguish bubbles from jet-like features and normal lobes, whose axial ratios are usually higher than $\sim\,$2. \cite{W78} suggests that below a critical luminosity, the outflowing plasma forms bubbles, whereas sources above this luminosity tend to form collimated radio jets. \cite{S83} also support the view that bubble formation is more likely to occur in low-luminosity systems. Non-thermal radio bubbles have been observed in other Seyferts (e.g. NGC\,1068, NGC\,2992, or NGC\,5548) and LINERs (NGC\,3079 or NGC\,3367) \citep{HS06}. According to this criterion, the NW-lobe of NGC\,4438 aligns with the bubble definition while the SE-lobe remain as a lobe.\\

Multi-wavelength approaches with similar high resolution and high sensitivity data at different frequencies, may shed light on the interpretation of these bubbles. Fig.\,\ref{RGB} allows us to appreciate the co-spatiality of the emitting features at the three analysed frequencies in the NW-lobe. The edges of the H$\alpha$ bubble, drawn by ionised gas, and the ridges of the main bubble that emits synchrotron radio emission are entirely coincident (see the white regions in Fig.\,\ref{RGB}). This complex northwestern radio structure observed and the bubble-like shape of the H$\alpha$ emission suggest a common nuclear power source, as already suggested by \citet{K02} \citep[see also][]{M11}. Often these bubbles show coincident X-ray emission (\citealt{C08}), as is also the case for the NGC\,4438 NW-lobe (Fig.\,\ref{RGB}).

Based on the sample selected by \cite{HS06}, they conclude that the formation of these bubbles is more closely linked to the presence of an AGN than with a starburst process. This scenario is also ruled out in our case due to the lack of an intense starburst in NGC\,4438 (\citealt{B05}). Instead, the bubble morphology is probably due to the jet interacting strongly with cold, dense material in the galaxy (e.g. \citealt{K10}). The co-spatial X-ray thermal emission observed would be also produced by the heating of gas by jet/ISM interaction, resulting in a greater mixing (model IIb in \citealt{C08}, more likely in clumpy ISM regions). This would resemble the concept of a `frustrated jet' discussed by \citep{Gui06}. Both observational studies and simulations indicate that the interaction between the jet and the ISM can cause significant perturbations in galactic morphology and kinematics \citep{M2017, C22, F22, M18, P24, M2025}. This impact could trigger the launch of outflows and the expansion or inflation of bubble-like structures (e.g. \citealt{M25} for the LINER NGC\,1167).

This hypothesis would result in the formation of H$\alpha$ shells, produced when cooler gas is compressed by energy injection from the jet. The clear spatial overlap between H$\alpha$ (green) and radio emission (red) in the NW-lobe of NGC\,4438 supports the hypothesis that a jet inflates the bubble while ionising the surrounding gas, which remains trapped around it.

X-rays, on the other hand, are found mainly within the bubble, showing hot thermal emission in the NGC\,4438 NW-lobe, as is the case for NGC\,6764 \citep{K10}. Temperatures found for this bubble (kT$\sim0.5\,keV$) are slightly lower than those for the bubble in NGC\,6764 \citep{C08}. The reason is most likely that they allow metallicity to vary, finding sub-solar values, whereas we fixed it to solar metallicity. Indeed, although the improvement in the spectral fit is not statistically significant, when solar metallicity is allowed to vary, higher temperatures are also found for the regions studied in this analysis. \citet{Li22} also use a non-thermal component to characterise the X-ray emission of the NGC\,4438 NW-lobe. However, a careful analysis of the significance of this component (see Sect.\,\ref{Results_xray}) shows that it is not statistically needed to explain X-ray emission of this bubble. Therefore, although a minor contribution from non-thermal emission cannot be ruled out, thermal emission dominates the X-ray emission observed in the NGC\,4438 NW-lobe.

The connection between radio and X-ray emission is naturally explained by jets emerging from AGN that can inject energy into the surrounding hot gas, inflating lobes or bubbles that displace the X-ray-emitting gas (\citealt{B04}; \citealt{F12}). This effect produces the so-called radio-filled X-ray cavities, i.e. depressions in the X-ray surface brightness that are often filled with radio-emitting plasma (e.g. \citealt{D08}). Therefore, the interaction between X-rays and radio emission in the central kiloparsec region is direct evidence of mechanical AGN feedback. In contrast to the X-ray cavities observed in clusters, we observe in the bottom panel in Fig.\,\ref{xray} an apparent central depression (hole) in synchrotron emission at radio maps that coincides spatially with enhanced X-ray emission. This structure may arise when a slightly collimated jet forms a cocoon that surrounds, rather than entirely displaces, the hot gas. It allows thermal plasma to remain or reaccumulate in the central region. This scenario is also compatible with the overall overlap between these emissions and the NW-lobe structure.

Interestingly, in NGC\,4438, X-ray emission also extends beyond the ionised shell. While projection effects may contribute, shock-heating of the ISM immediately increases its temperature when the ISM is sufficiently dense and/or inhomogeneous (\citealt{Gui06}). Thus, the pre-shock gas can be heated directly before the H{$\alpha$} shell forms and X-ray emission could be present beyond it.
 
A part of the co-spatial emission, NGC\,4438 also shows features that do not have a direct link at other wavelengths. In Fig. \ref{RGB}, we show extended H$\alpha$ emission perpendicular to the axis of the double-lobe morphology, that is not related to the AGN. Instead, it could be ionised gas originating from young stars. Spectroscopic evidence shows the expected rotational motion in the galaxy \citep{HM24} and non-AGN ionisation hints in this region \citep{L25}. Finally, an X-ray point-like source is seen to the south-west from the nucleus. It is consistent with a non-thermal origin (fitted to an absorbed power-law, see Sect. \,\ref{Results_xray}), with luminosity consistent with being an X-ray binary system.

\subsection{Energetics}\label{diss_energetic}

We present an energetic model that traces structures at scale of $\sim$ 0.4\,arcsec, following the minimum energy assumption (see Sect. \ref{Results_maps}). We obtain maps for different parameters including the $B_{min}$ or the $t_{syn}$ (see Fig. \ref{fig:energymaps} and Table \ref{table:4}). The sub-arcsecond resolution in this analysis allows us to treat each substructure individually. In particular, this enables the AGN to be distinguished from the lobes, which allow us to interpret the physics in the central kiloparsec thoroughly. The total radio power taking into account the three main radio structures (the main bubble, the SE-lobe and the nucleus, see Sect. 5.5) is $\sim 1.22 \times 10^{40}\,$ erg\,s$^{-1}$. Our model covers from 4 to 8\,GHz, following the model that first proposed \cite{H07}. In fact, they provide this model at scales of 1.5\,arcsec and at lower frequencies compared to our work, between 1.49 and 4.86 GHz. In general, both models are consistent, but the total radio power reported by \cite{H07} is a few times $10^{42}\,$ erg\,s$^{-1}$, two orders of magnitude higher than in this work. Nevertheless, after a revision and recalculation, the total radio power they estimated is $ 1.7 \times 10^{40}\,$ erg\,s$^{-1}$ (private communication with Professor J. Irwin and Dr. A. Hota), which is compatible with our result. 

The mean values of the energetic parameters (Fig. \ref{fig:energymaps} and Table \ref{table:3} in this paper, Fig. 3 and Table 3 in \citealt{H07}) highlight the differences between these two works: 1) the higher resolution of our images, gives us a more accurate and detailed results, 2) the higher sensitivity of our analysis means that when we apply the S/N = 5 cut-off, we avoid regions with higher error, such as the borders and diffuse emission between lobes, that are included in \cite{H07} affecting the mean values.

We calculate the jet power applying Eq. 8 that relates the radio power and the kinetic power of a radio core. We obtain a jet power between $\sim$ 10$^{43}$ and 10$^{44}$ \,erg\,s$^{-1}$, depending on the substructures selected (see Sect. \ref{diss_connection}). Comparable LINERs, such as NGC\,4374, NGC\,4636, or NGC\,5846 show jet powers of $\sim$ 10$^{42}$\,erg\,s$^{-1}$ \citep{M07}. Both NGC\,2911 and NGC\,1369, which are included in \cite{M14}, exhibit jets extending  out to $\sim$\,500\,pc, resembling those of NGC\,4438. Their jet powers are 5.9$\times$ 10$^{42}$\,erg\,s$^{-1}$ and 1.5 $\times$ 10$^{42}$\,erg\,s$^{-1}$, respectively. Even the prototypical LINER NGC\,1052, which has an outflow of 8.8 $\pm$3.5$\times$ 10$^{40}$\,erg\,s$^{-1}$ and a 3 kiloparsec jet, does not show such a powerful jet \,(6$\times$10$^{42}$\,erg\,s$^{-1}$). Apparently, the jet power of NGC\,4438 is higher than that of comparable objects. This is consistent with the high mass outflow rate (1.48$\pm$0.99  M$_\odot$ yr$^{-1}$) found in \cite{HM24}.

\begin{figure}[h]
       
        \begin{subfigure}[t]{0.4\textwidth}
        \hspace*{0.25\textwidth}
            \includegraphics[width=\linewidth]{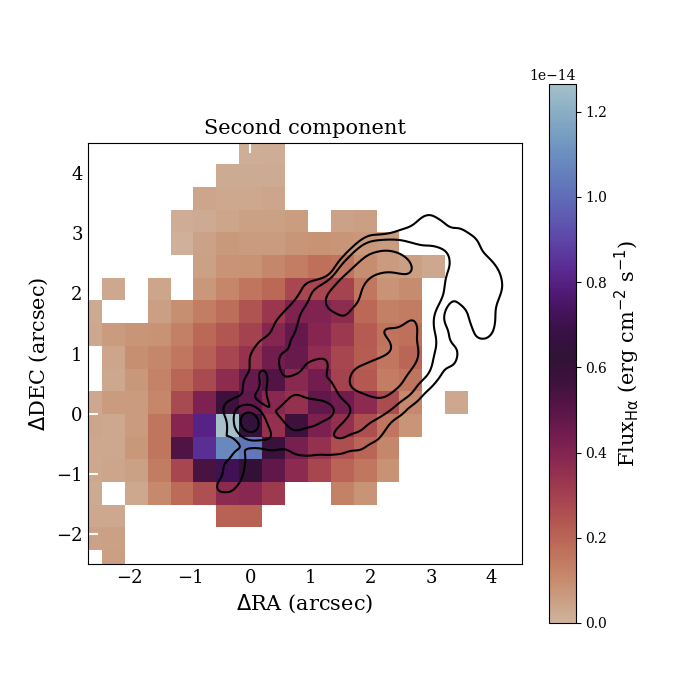}
            
        \end{subfigure}%
        
        \begin{subfigure}[t]{0.4\textwidth}
        \hspace*{0.25\textwidth}
            \includegraphics[width=\linewidth]{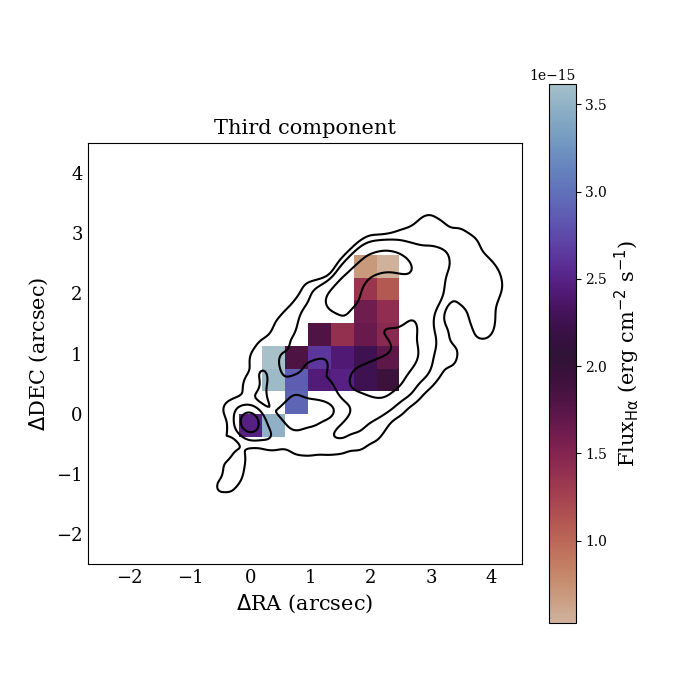}       
        \end{subfigure}
        \caption{Flux maps of ionised gas (H$\alpha$) kinematical components of NGC4438 with C-band radio contours (0.08 mJy, 0.2mJ and 0.35 mJy). The  maps show the integrated H$\alpha$ flux per spaxel of MEGARA, and were provided by \cite{HM24}. Both components are related to non-rotational motions (i.e outflow). See all the details in \cite{HM24}.}
        \label{components}
\end{figure}

\subsection{Connection between the outflow and the radio bubble} \label{diss_connection}

We combine the radio analysis from this study with the kinematical IFS analysis presented by \cite{HM24}, (see Fig. \ref{components}), to discuss the potential connection between the ionised gas outflow and the radio bubble. In Fig. \ref{components}, we show the flux maps of the second and third H$\alpha$ kinematical components (see above) with C-band radio contours overlapped. The spaxels with more flux in the second component trace the NW-lobe, showing hints of the conical morphology in the north-west direction. The third component fits perfectly the main bubble location in the NW-lobe, corresponding to the more compact radio emission. Both co-locations are consistent with a scenario where the ionised gas and the soft X-ray emission arise in regions where the bubbles of synchrotron-emitting plasma heat and interact with the surrounding ISM.

\cite{HM24} studied the kinematic properties of nine LINERS, including NGC\,4438, with IFS data (GTC/MEGARA) in order to explore the extent of their ionised gas outflows. They fitted the emission lines (i.e. H$\alpha$, [SII] and [OIII]) using a multi-Gaussian fit decomposition technique and identified ionised gas outflows as a secondary, blue-shifted, broad component, that could not be explained by rotational motions (see also \citealt{C18} for details on the fitting procedure). For NGC\,4438, they found three components: while the primary component traces the gas disc (rotational motion), the secondary and the third components (Fig. \ref{components}) are most likely related to the detected non-rotational motions and the H$\alpha$ bubble.  
\cite{HM24} identified an ionised gas outflow extending 307 pc (projected distance) in the north-west direction, with a kinetic power of $ 9.9 \pm 7.9  \times 10^{40}$\,erg\,s$^{-1}$, based on the whole second component extension due to its co-spatiality with the H$\alpha$ bubble seen in HST images (\citealt{M11}). However, \cite{HM24} did not take into account the third component in the physical characterisation of the outflow since they did not find evidence of its origin and outflow relation. In contrast, in Fig. \ref{components} (bottom panel), we show a clear morphological overlap between the third component and the radio emission, not only in the direction but also in the extension up to the main bubble. We evaluated the specific contribution of the third H$\alpha$ kinematical component to the ionised outflow mass to determine if this kinematical component of H$\alpha$ could be related and significant to the total ionised outflow mass and energetics. Using the data already published in \cite{HM24}, and following the same method, we find this component has a higher electron density (computed from the ratio between the lines of [SII] doublet) and an outflow mass of $\sim$ 2.3$\times$10$^{4}$ M$_\odot$, one order of magnitude lower than that calculated using the second component. Thus, although we associate the third component with the first jet-ISM interaction, adding the third component contribution to the total power of the outflow does not imply any significant variance. Moreover, we also attempted to mask the nuclear emission; however, the final power that does not show important variations. Overall, we consider the power of the outflow given by \cite{HM24} as the reference value. 

The co-spatial radio emission that we have detected is extended out to 400 pc from the nucleus to the hook. The high resolution of our energetic model allows us to discern the potential impact of the different substructures on the kinematics of ionised gas. So far, we have calculated the kinetic (jet) power (Eq. 8) by taking into account the different substructures. Considering the whole radio emission within the second H$\alpha$ component, so including the nucleus and the entire NW-lobe (base of the cone, main bubble and extended hook), the radio power is 3.31$\times$10$^{40}$ erg\,s$^{-1}$ and the jet power 5.26$\times$10$^{44}$ erg\,s$^{-1}$. If we consider only the nucleus as the radio core with a non-resolved jet, its radio power is 6.7$\times$10$^{38}$ erg\,s$^{-1}$ and the mean jet power would be 2.23$\times$10$^{43}$ erg\,s$^{-1}$. For the NW-lobe without the nucleus, the radio power is 3.24$\times$10$^{40}$ erg\,s$^{-1}$ and the jet power 5.17$\times$10$^{44}$ erg\,s$^{-1}$. This indicates that the NW-lobe contribution is more significant than the compact AGN contribution, which would account for less than 5$\%$. Nevertheless, the power of a potential nuclear jet would be enough to launch the observed outflow (see above). In the NW-lobe, the main bubble represents around 30$\%$ of its kinetic power. The base of the cone increases the final value considerably, representing around 70$\%$, but the error in this region could be higher. 
In any case, based on an energetic analysis and a comparison of outflow and jet powers, we conclude that the jet-driven scenario is fully compatible, regardless of whether the different substructures are combined or treated separately. In this analysis, we note that the outflow definition includes not only the bubble, but also the displaced ionised gas. Anyway, even when the radio emission comes from the very nuclear part, the hypothesis remains consistent; the radio power from the AGN is sufficient to disturb and displace the ionised gas (see Fig. \ref{components}). Due to the assumptions made in Sect. \ref{Data}, the error associated with jet power could be significant. Taking into account only the errors coming from the flux and the spectral index, the error could reach 30\%. Note also that, our model dates the oldest structures to be around 2 Myr old, which agrees with the 1.4\,Myr that can be derived from the results in \cite{HM24}.
\newline
\newline
\indent NGC\,4438 shows similarities with other LINERs and Seyferts, but there is insufficient evidence to establish whether they share the same scenario in terms of feedback mechanisms. NGC\,1052, the prototypical LINER, was analysed by \cite{C22}. They also identify three kinematical components for NGC\,1052 with IFS MUSE data and also suggest that the ionised gas outflow is most likely driven by the radio jet due to its coincident direction and energetic compatibility. In contrast to NGC\,4438, NGC\,1052 is an isolated elliptical galaxy, and its X-ray and H{$\alpha$} morphologies are relatively symmetric. The Seyfert/LINER spiral galaxy NGC6764 shows a reported relation between the radio and X-ray bubbles (\citealt{C08};\citealt{K10}) similar to NGC4438, although this galaxy has a relevant starburst contribution. Another interesting  target is the nearby Seyfert 2, NGC\,1068; similar to NGC 4438, it is a galaxy with an asymmetrical double-lobed jet structure and bow shocks (\citealt{M24}), similar to the ridges or shells we have referred to in this work. The dense medium in the centre of the galaxy also plays a significant role in its morphology, even causing the jet to change direction (\citealt{G04};\citealt{M24}). Comparing the energies of the molecular outflow (\citealt{GB14}) and the jet also supports the jet-driven scenario in NGC\,1068 (\citealt{M24}). Despite all the similarities, the collimated jet is much more clearly detected NGC\,1068. Moreover, this isolated barred spiral galaxy has a jet oriented along the starburst bar (\citealt{GB14}). In order to better understand LINERs, a systematic multi-frequency study with a significant number of sources is required. It will allow us to draw more comprehensive conclusions and identify trends in this particular type of LLAGNs.

\section{Summary and Conclusions}\label{conclusions} 

This work proposes a multi-wavelength approach to investigate the feedback in a low-accretion, radiatively inefficient AGN, the LINER NGC\,4438. The wavelengths involved in the study are the optical (H${\alpha}$ imaging and integral field spectroscopy), radio (data from e-MERLIN and VLA), and \textit{Chandra} X-ray data. 
Firstly, we calibrated and imaged the radio (L, C and X-band) and X-ray data. We used the C-band image to compute and analyse the spectral index. We confirm the detection of non-thermal synchrotron emission, which is consistent with the presence of a jet. We then searched for evidence of the jet-mode feedback through the multi-wavelength morphologies and estimating the energetics of the radio emission. Our main goal was to find out whether the radio jet is powerful enough to drive the ionised gas outflow. We present a model that estimates different parameters such as the age of the structures and their power.

NGC\,4438 shows a particular asymmetric central morphology, both in H${\alpha}$ and radio images. We distinguish three main radio-structures, namely the NW-lobe, the nucleus and the SE-lobe. The kinematical characterisation of the ionised gas outflow of NGC\,4438 by \cite{HM24} in their IFS study, allows us to compare the energetics of ionised gas outflow with that of radio and X-ray emissions reported in this work. The study presented here leads to the following conclusions:

   \begin{enumerate}
      \item We have analysed the high-resolution radio emission of NGC\,4438, which enables us to reach sub-kiloparsec scales. The study reveals the nuclear features and different substructures within the NW-lobe. The spectral index analysis shows $\alpha\,\sim\,-0.8$ in NW-lobe and $\alpha$\,$\sim$\,$-0.4$ in the SE-lobe, whereas it is flat for the nucleus ($\alpha$\,$\sim$\,0).  
      \item The multi-wavelength images presented in this paper show clear connections between emissions, demonstrating a co-spatial relationship between the ionised gas outflow, the jet and the X-ray emission. 
      \item The energetic study at radio frequencies shows a compatibility between the energetics of the ionised outflow and the radio jet. The power of the jet ($\sim\,5\times 10^{44}$ erg\,s$^{-1}$), is about three orders of magnitude the ionised gas outflow power ($ 9.9   \times 10^{40}\,$ erg\,s$^{-1}$), so consistent with the jet-driven scenario. 
      \item Thermal emission dominates at X-ray in the NW-lobe, although a minor contribution from non-thermal emission cannot be ruled out. We observe how the jet surrounds the hot gas (X-ray), allowing the thermal plasma to remain or reaccumulate in a radio cavity. We only find evidence on non-thermal emission in the nuclear region. 

   \end{enumerate}

NGC\,4438 is the first source that we have analysed, and it provides a pilot for a larger sample. In fact, we are pursuing a systematic study applying this methodology to a sample of 30 LINERs, from the H$\alpha$ imaging atlas presented in \cite{HM22}, for which we have radio, X-ray and IFU data. The sample is unbiased in terms of both host-galaxy morphology and nuclear radio morphology. Our aim is to evaluate the role of the jet, if present, whether resolved or unresolved, collimated or disrupted. We seek to determine whether jet-mode activity is linked to the observed ionised gas outflows. High-resolution data across different wavelengths are crucial for these studies, such as those provided by VLA or e-MERLIN in radio, or sensitive optical IFS instruments as MUSE/VLT or MEGARA/GTC.

\begin{appendix}
\section{Acknowledgements}
Authors acknowledge the anonymous referee for their constructive comments to improve the paper.

We acknowledge the VLA operations and support team for their assistance in the preparation and execution of the X-band 24B-323 project observations. The VLA is operated by the National Radio Astronomy Observatory (NRAO). The NRAO is a facility of the National Science Foundation operated under a cooperative agreement by Associated Universities, Inc.  We thank the e-MERLIN Observation Support Scientists for the successful acquisition of the data and for their accuracy and dedication in ensuring proper calibration of L-band CY16021 data. e-MERLIN is a National Facility operated by the University of Manchester at Jodrell Bank Observatory on behalf of Science and Technologies Facilities Council (STFC).\\

We thank the attention and feedback from Professor Judith Irwin and Dr. Ananda Hota, regarding their model application.\\

We thank Mr. Lucatelli for the interesting discussions about the methodology presented. \\

MPS, IM, JMasegosa, SC acknowledge financial support from the Spanish Ministerio de Ciencia, Innovación y Universidades (MCIU)  for the Spanish MCIU grant PID2022-140871NB-C21. MPS acknowledges financial support from MCIU under the grant PRE2021-100265, which is part of the SEV-2017-0709.  \\

JMoldón acknowledges financial support from the National grant PID2023-147883NB-C21, funded by MCIU/AEI/ 10.13039/501100011033. \\

MPS and JMoldón acknowledge the Spanish Prototype of an SRC (SPSRC) service and support funded by the MICU, by the Junta de Andalucía, by the European Regional Development Funds (ERDF) and by the European Union NextGenerationEU/PRTR. The SPSRC and MPS, JMoldón, IM, JMasegosa, SC and OGM  acknowledge financial support from the Agencia Estatal de Investigación (AEI) through the `Center of Excellence Severo Ochoa to the Instituto de Astrofísica de Andalucía (IAA-CSIC) (SEV-2017-0709) and CEX2021-001131-S funded by MICIU/AEI/10.13039/501100011033.\\

OGM acknowledges the financial support from the Ciencia de Fontera project CF-2023-6100 (SECIHTI) and PAPII project IN109123 (UNAM). \\
LHM acknowledges financial support by the grant PID2021-124665NB-I00 funded by the Spanish Ministry of Science and Innovation and the State Agency of Research  MCIN/AEI/10.13039/501100011033 PID2021-124665NB-I00 and ERDF A way of making Europe. \\

MPS has received funding from the European Union’s Horizon 2020 research and innovation programme under grant agreement No 101004719 (Opticon Radionet Pilot, ORP), employed in a e-MERLIN data training.

\end{appendix}

\bibliography{references}

\end{document}